\documentclass[11pt]{article}
\usepackage{jheppub}

\usepackage{epsfig}
\usepackage{amssymb}
\usepackage{amsmath}
\usepackage{amsfonts}

\newcommand{\be}{\begin{equation}}
\newcommand{\ee}{\end{equation}}
\newcommand{\bea}{\begin{eqnarray}}
\newcommand{\eea}{\end{eqnarray}}
\newcommand{\bi}{\begin{itemize}}
\newcommand{\ei}{\end{itemize}}
\newcommand{\ben}{\begin{enumerate}}
\newcommand{\een}{\end{enumerate}}

\newcommand{\lp}{\left(}
\newcommand{\rp}{\right)}
\newcommand{\ep}{\epsilon)}

\newcommand{\nn}{\nonumber}

\def\gsim{\mathrel{\rlap{\lower4pt\hbox{\hskip1pt$\sim$}}
    \raise1pt\hbox{$>$}}}         
\def\lsim{\mathrel{\rlap{\lower4pt\hbox{\hskip1pt$\sim$}}
    \raise1pt\hbox{$<$}}}         

\def \as {\alpha_s}
\def \ep {\epsilon}

\begin{document}


\preprint{ANL-HEP-PR-13-13}

\title{Higgs boson production in association 
with a jet at next-to-next-to-leading 
order in perturbative QCD}

\author[1]{Radja Boughezal,}
\author[2]{Fabrizio Caola,}
\author[2]{Kirill Melnikov,}
\author[1,3]{Frank Petriello}
\author[1]{and Markus Schulze}

\affiliation[1]{High Energy Physics Division, Argonne National Laboratory, Argonne, IL 60439, USA}

\affiliation[2]{Department of Physics and Astronomy, Johns Hopkins University, Baltimore, MD 21218, USA}

\affiliation[3]{Department of Physics \& Astronomy, Northwestern University, Evanston, IL 60208, USA}

\emailAdd{rboughezal@anl.gov}
\emailAdd{caola@pha.jhu.edu}
\emailAdd{melnikov@pha.jhu.edu}
\emailAdd{f-petriello@northwestern.edu}
\emailAdd{markus.schulze@anl.gov}

\abstract{We report on  a calculation of the 
cross-section 
for Higgs boson production in gluon fusion 
in association with a hadronic jet at next-to-next-to-leading 
order (NNLO) 
in perturbative QCD. The computational technique is discussed in detail. 
We show explicitly how to employ known 
soft and collinear limits of scattering 
amplitudes to construct subtraction terms for NNLO computations. 
Cancellation of singularities is demonstrated numerically 
for the collinearly-subtracted $gg \to H + j $  cross-section through 
NNLO and the finite $\sigma_{gg \to Hj}$ cross-section 
is computed 
through  ${\cal O}(\alpha_s^5)$ 
as a function of the center-of-mass collision energy.  We present numerical results for the gluon-fusion contribution to 
Higgs production in association with a jet at the LHC.  The NNLO QCD corrections significantly reduce the 
residual scale dependence of the cross-section.
The computational  method that we describe in this paper 
is applicable to the calculation  of NNLO QCD corrections 
to any other $ 2 \to 2$ process at a hadron collider  
without modification.
}

\maketitle

\section{Introduction} 

The ATLAS and CMS
experiments at the Large Hadron Collider (LHC) have discovered a 
new particle with a mass of approximately $125~{\rm GeV}$
\cite{discovery-atlas,discovery-cms}  whose properties 
are consistent with that of the Standard Model Higgs boson. 
Continuing  studies at the LHC are  focusing  on the detailed 
understanding   of the quantum numbers of this particle \cite{Chatrchyan:2012jja}
and its  couplings  to gauge 
bosons and fermions (see e.g. \cite{LHCHiggsCrossSectionWorkingGroup:2012nn} and references 
therein).  The successful completion of this 
 task is crucial 
for determining   if the new particle  is indeed 
the long-awaited Higgs boson  or instead some other state.

A reliable understanding of coupling constants cannot occur without 
accurate theoretical predictions for the main Higgs boson production 
and decay processes.  Arriving at such predictions requires 
the computation of higher-order QCD corrections, since they are known 
to affect Higgs production rates and decay branching fractions 
in a significant way. In fact, for gluon-initiated processes 
$gg \to H + X$,  where $X$ is a state with zero, one or two 
hard jets, the next-to-leading order (NLO)
QCD radiative corrections are known to be so large 
\cite{dawson,djouadi,deflorian,
ravindran,glosser,jmc1,jmc2,
vanDeurzen:2013rv}
that 
next-to-next-to-leading order (NNLO) QCD computations are important 
for reliable phenomenology.

Unfortunately, current computational technology only allows 
NNLO QCD computations for the case of Higgs boson production 
in association with zero jets \cite{Harlander:2002wh,kb,Ravindran:2003um,babis,grazi}. 
Extending this result to one or more 
jets will lead to a refined  analysis of the 
$pp \to H \to W^+W^-$  process,  since in that case 
final states  with different  jet multiplicities  
are treated as different processes in order to optimize search strategies. 
The information about  the relative significance of the Higgs boson production 
in association with zero, one or two jets is currently extracted from available 
fixed-order computations supplemented with  resummations  of 
the most important terms in the perturbative expansion~\cite{salam, Becher:2012qa, Tackmann:2012bt,Liu:2012sz}.  Explicit NNLO computations for multi-jet processes  
will  be indispensable for understanding the reliability of 
these predictions. 

The other motivation for  this work is of a more theoretical 
nature.  One can argue that  the framework of perturbative QCD that has been developed since the late 1970's has proven to be 
one of the most important areas of particle physics phenomenology.
Indeed, it is {\it impossible} to imagine contemporary high-energy 
physics without hadron collider physics whose proper  
description is intimately related with parton shower event generators, 
sophisticated fits of parton distribution functions, fixed-order 
perturbative calculations and the like.   Much of our understanding 
of perturbative QCD is based on how soft and collinear singularities 
cancel in suitable ``infra-red safe'' quantities, since this defines 
short-distance observables that can be calculated in perturbation 
theory.  Currently, there exists an interesting gap in this understanding. On one hand, 
general theorems~\cite{Kinoshita:1962ur,Lee:1964is} 
ensure that this cancellation occurs in suitably 
defined quantities to all orders in perturbation theory. 
On the other 
hand, we {\it only} know  how to use 
those ideas for generic computations
of infra-red-safe observables at leading and next-to-leading order in 
perturbative QCD 
\cite{Catani:1996vz,Frixione:1995ms}.  It is still not entirely clear how to construct a 
general  calculational  scheme for two- and higher-loop 
 computations.  

It is important to point out that, in spite of the fact that a 
generic computational scheme is not available, 
a fairly large number of NNLO computations 
for various processes have already been performed 
\cite{grazi,m1,Melnikov:2006di,cat2,an2,babis,an4,g1,g2,Weinzierl:2008iv,Anastasiou:2005pn,
Biswas:2009rb,Melnikov:2008qs,Ferrera:2011bk,Anastasiou:2011qx,Catani:2011qz}, but 
until very recently such computations always utilized 
a particular aspect  of a specific process. 
Such aspects included a small number 
of final-state particles, or their color neutrality, 
or absence of color-charged 
particles in the initial state, or even the fact that all matter 
particles in a particular process were massive.  A generic 
algorithm that is valid {\it irrespective} of the details of the 
process under consideration was not  worked  out.  This  situation is somewhat peculiar, because following 
the successful development of generic methods for NLO computations~\cite{Catani:1996vz,Frixione:1995ms}, 
it was generally felt that the development of similar methods 
for NNLO computations would be relatively straightforward. 
For this reason, about ten years ago  
many authors calculated  infra-red and collinear 
limits of generic QCD amplitudes 
\cite{lcat1,lb,lmc3,lcat2,lzb1,lcat3,lcat4,Kosower:1999rx}
that are potentially relevant for NNLO  computations, and 
a large number of two-loop 
$2 \to 2$ scattering  amplitudes became available \cite{twol,3jme}. 
Unfortunately, 
since it  proved harder than expected to develop a working 
scheme for NNLO computations, these 
infra-red and collinear  limits were never 
used for their intended purpose. 

Recently, important steps towards developing a general  computational scheme 
valid through NNLO QCD that, at least in principle, 
is applicable to 
processes of arbitrary multiplicity, 
were made  by Czakon~\cite{Czakon:2010td,Czakon:2011ve},
who suggested to combine the ideas of sector decomposition 
\cite{binothheinrich1,binothheinrich2,an1}
and Frixione-Kunszt-Signer (FKS) \cite{Frixione:1995ms} 
phase-space partitioning.
These results were used to obtain  NNLO QCD 
corrections to the cross-sections for  $q \bar q \to t \bar t$ 
\cite{Baernreuther:2012ws,Czakon:2012zr} 
and $qg \to t \bar t$ \cite{Czakon:2012pz} processes.
A similar computational scheme was also 
developed  in Ref.~\cite{Boughezal:2011jf} where it was applied 
to the calculation of  NNLO  QED corrections  to $Z \to e^+e^-$. 
We note that parallel developments in the antennae subtraction 
technique~\cite{antennae} have recently led to the calculation 
of the NNLO QCD corrections to the leading-color 
all-gluon contribution to di-jet production at the LHC~\cite{Ridder:2013mf}.

While the above results mark important progress in developing a suitable framework for NNLO computations, 
a large number of technical aspects 
still needs to be understood and worked out.  
It is best to 
do this by considering a realistic example with 
all the ensuing complications. This is 
the purpose of the current paper.  We consider 
the hadro-production of the Higgs boson in association  
with one hadronic  jet at NNLO in QCD. 
To make calculations as simple as possible, we work in pure gluodynamics, 
i.e. QCD without light fermions.  We note that an understanding of how to compute 
the NNLO QCD corrections  to $H+j$ production 
is instructive because this process possesses  all  non-trivial 
aspects of a generic NNLO QCD problem. Indeed, 
\begin{itemize} 
\item   there are  colored particles in the  initial state; 
\item  there are colored particles in the final state; 
\item  already at leading order, 
the total  cross-section for this process does not exist unless a
jet algorithm  is specified; 
\item  this process exhibits 
the most general structure  
of infra-red and collinear singularities, 
since these singularities occur due 
to radiation of gluons in the initial and final 
states; 
\item singular collinear splittings $g \to gg$  
involve non-trivial  spin correlations; 
\item the number of Feynman diagrams that we need to compute 
is large.\footnote{For example, 
the process  $gg \to Hggg$  is described at leading order  
by 230 diagrams while the $gg \to Hgg$ process at one-loop 
is described by 603 diagrams.} 
\end{itemize} 
The only ``non-generic''
feature of the process that we consider 
is the high symmetry of the final state which simplifies the bookkeeping 
and speeds up the computation. We feel, however, that having 
this simplification  is useful in  the first step in the development 
 of the new technology and that 
it does not affect the generality  of the method
that we describe  in this paper. 

The remainder of the paper is organized 
as follows. In the next Section we describe the setup of the 
calculation.  In Section~\ref{phasespace} we discuss the parametrizations 
of the phase-space for leading, next-to-leading and next-to-next-to-leading 
order computations. In Section~\ref{singularlimits} we explain how singular 
limits of amplitudes are used. 
In Section~\ref{ampl_ep} we describe how ${\cal O}(\ep)$ parts 
of the relevant amplitudes can be computed using helicity methods. In Section~\ref{sec:num} 
we describe the numerical implementation of our method. In Section~\ref{sec:checks} 
we discuss some tests and 
show the results of the computation. We conclude in Section~\ref{sec:conclusions}.
Some useful formulae are given in the Appendix.

\section{The setup}
\label{setup}

We are interested in the computation of NNLO 
QCD corrections to the process $g + g \to 
H+g $, where the Higgs boson can decay into arbitrary  
particles.  To compute this and related processes, we use the QCD 
Lagrangian, supplemented with a dimension-five non-renormalizable 
operator that describes the interaction of the Higgs boson with gluons 
in the limit of very large top quark mass
\be
{\cal L} = -\frac{1}{4} G_{\mu \nu}^{(a)} G^{(a),\mu, \nu}
- \lambda_{Hgg} H G_{\mu \nu}^{(a)} G^{(a),\mu, \nu}.
\label{eq_lag1}
\ee
Here, $G_{\mu \nu}^{(a)}$ is the field-strength tensor 
of the gluon field and $H$ is the Higgs boson field. 

Matrix elements computed  with the Lagrangian of Eq.~(\ref{eq_lag1}) 
need to be renormalized; to do so, two renormalization constants 
are required.  The first one relates 
bare and renormalized QCD coupling constants
\be
\alpha_s^{(0)}\mu_0^{2\ep} S_\ep = \mu^{2\epsilon} \alpha_s Z_{\alpha_s},
\;\;\
Z_{\alpha_s} =
1 - \frac{\beta_0}{\ep} 
\left ( \frac{\alpha_s}{2\pi} \right )  
+ \left ( \frac{\beta_0^2}{\ep^2}
-\frac{\beta_1}{2\ep} \right ) \left ( \frac{\alpha_s}{2\pi} \right )^2
+....
\ee
Here, $\alpha_s$ is the strong 
coupling constant in the ${\overline {\rm MS}}$ scheme  evaluated 
at the renormalization scale $\mu$,
$
S_\ep = (4\pi)^{-\ep} e^{-\gamma \ep},\;
\gamma = 0.5772$
is the Euler constant
and
\be
\beta_0=\frac{11 N_c}{6},
\;\;\;\;\;\;\;\beta_1=\frac{17N_c^2}{6}
\label{eq_beta} 
\ee
are one- and two-loop contributions to the QCD $\beta$-function
computed with the  Lagrangian of Eq.~(\ref{eq_lag1}). $N_c=3$ is the 
number of colors.
We note that Eq.~(\ref{eq_beta}) is only valid in a theory without light 
fermions, as defined  
by the Lagrangian Eq.~(\ref{eq_lag1}).

The second renormalization constant ensures that matrix elements 
of the $HGG$ dimension-five operator are finite. 
It reads 
\be\label{lambdadef}
\lambda_{Hgg}^{(0)} = 
-\frac{\alpha_s}{12 \pi v} C(\alpha_s) Z_{\rm eff}(\alpha_s),
\;\;\;\
Z_{\rm eff} = 
1 - \frac{\beta_0}{\ep} 
\left ( \frac{\alpha_s}{2\pi} \right )  
+ \left ( \frac{\beta_0^2}{\ep^2}
-\frac{\beta_1}{\ep} \right ) \left ( \frac{\alpha_s}{2\pi} \right )^2
+..
\ee
In the above formula, $C(\alpha_s)$ is the Wilson coefficient 
of the  $H G G$ operator    
in the ${\overline {\rm MS}}$ scheme \cite{kostya}
\be
C = 1 + \frac{11}{2} \left ( \frac{\alpha_s}{2\pi} \right )
+ \left ( \frac{\alpha_s}{2\pi}   \right )^2 
 \left [ \frac{2777}{72} + \frac{19}{4} \ln \frac{\mu^2}{m_t^2} \right ] 
+ {\cal O}(\alpha_s,N_f), 
\ee
where $m_t$ is the mass of the top quark.  We 
emphasize again that the displayed  result is only valid in the approximation 
when no light fermions are present in the theory. 

Renormalization of the strong coupling constant and of the effective 
Higgs-gluon coupling removes ultraviolet divergences from the matrix 
elements. The remaining divergences are of infra-red origin. To 
remove them, we must both define and compute infra-red safe observables, 
and absorb remaining collinear singularities by renormalizing 
parton distribution functions. We now discuss these two issues.

Generic infra-red safe observables are defined using jet algorithms. 
For the calculation described in this paper  
we employ  the $k_\perp$-algorithm. This algorithm belongs to the class 
of sequential jet algorithms.  It requires specification of 
the minimal  transverse momentum of the reconstructed jets $p_{\perp,j}$ 
and 
the minimal ``angular'' distance between two partons, 
$\Delta R_{\ij} = \sqrt{(y_i-y_j)^2 + (\varphi_i- \varphi_j)^2}$, 
where $y = 1/2 \ln (E+p_z)/(E-p_z)$ is the rapidity and $\varphi$ 
is the azimuthal 
angle of a parton. Once 
$\Delta R$ and $p_{\perp,j}$  are specified, 
the jet algorithm maps a set of parton momenta onto a set of jet momenta in such a way 
that jet momenta are stable against soft and collinear parton splittings. 
The Kinoshita-Lee-Naunberg  theorem \cite{Kinoshita:1962ur,Lee:1964is} 
then ensures that observables constructed from 
jet four-momenta are determined by short-distance physics and can therefore be computed in QCD perturbation theory.  
However, because massless colored partons are present in the initial state of the partonic 
process $gg \to H + X$, the infra-red and collinear cancellation is not complete, 
even in the presence 
of a jet algorithm.  Collinear singularities associated with 
gluon radiation by incoming partons must be removed by additional renormalization of parton 
distribution functions.  We describe how to perform this renormalization in what follows. 
For definiteness, we  
focus our discussion 
on the  production cross-section 
of a Higgs boson and a jet in pure gluodynamics.

We denote the UV-renormalized 
partonic cross-section for the production of the  Higgs boson 
and a jet in a gluon fusion  by ${\bar \sigma}(x_1,x_2)$, and the collinear-renormalized 
partonic cross-section by  $\sigma(x_1,x_2)$. Once we know $\sigma(x_1,x_2)$, we 
can compute the hadronic cross-sections by integrating 
the product of $\sigma$  and 
gluon  distribution  functions  over $x_1$ and $x_2$
\be
\sigma(p +p \to H+j) = \int {\rm d} x_1 {\rm d}x_2 \;g(x_1) g(x_2) \; \sigma(x_1,x_2).
\ee
The relation between $\sigma$ and ${\bar \sigma}$ is given by the following formula
\footnote{We show 
this relation for pure gluodynamics; if more species of partons are 
present, Eq.~(\ref{eq_basic})  becomes 
a matrix equation.}
\be
\sigma = \Gamma^{-1} \otimes {\bar \sigma} \otimes \Gamma^{-1},
\label{eq_basic}
\ee
where the convolution sign stands for 
\be
\left [ f \otimes g \right](x) = 
\int \limits_{0}^{1} {\rm d} x {\rm d} y \delta( x - yz) f(y) g(z).
\ee
The collinear counter-terms are defined as 
\be
\Gamma = \delta(1-x) - \left ( \frac{\alpha_s}{2\pi} \right ) \Gamma_1 
+ \left ( \frac{\alpha_s}{2\pi} \right )^2  \Gamma_2,
\ee
with
\be
\Gamma_1 = \frac{P_{gg}^{(0)}}{\ep},\;\;\;\;
\Gamma_2 = \frac{1}{2\ep^2} \left ( P_{gg}^{(0)} \otimes P_{gg}^{(0)} + \beta_0 P^{(0)}_{gg} \right ) 
 - \frac{1}{2\epsilon} P_{gg}^{(1)}.
\ee
The relevant splitting functions and their convolutions
are given in the Appendix.  We write the UV-renormalized partonic cross-section through NNLO as  
\be
{\bar \sigma} = {\bar \sigma}^{(0)} 
+
\left ( \frac{\Gamma(1+\ep) \alpha_s}{2\pi} \right ) 
 {\bar \sigma}^{(1)} 
+ \left (\frac{ \Gamma(1+\ep)  \alpha_s}{2\pi} \right )^2 {\bar \sigma}^{(2)}, 
\ee
and the collinear-renormalized partonic cross-section as 
\be
 \sigma = \sigma^{(0)} + 
\left ( \frac{ \alpha_s}{2\pi} \right )   {\sigma}^{(1)} 
+ \left ( \frac{ \alpha_s}{2\pi} \right )^2 {\sigma}^{(2)}. 
\ee
We note that  the collinear-renormalized cross-section is finite. 
We then use Eq.~(\ref{eq_basic}) to obtain 
\be
\begin{split} 
& 
\sigma^{(0)} = {\bar \sigma}^{(0)},\;\;\;\;\;\;\;\;\;\;\;
\sigma^{(1)} = {\bar \sigma}^{(1)}
 + \frac{\Gamma_1 \otimes {\sigma}^{(0)}}{\Gamma(1+\ep)} 
 + \frac{{\sigma}^{(0)} \otimes \Gamma_1 }{\Gamma(1+\ep)},
\\
& \sigma^{(2)} = {\bar \sigma}^{(2)} 
- \frac{\Gamma_2 \otimes {\sigma}^{(0)} }{\Gamma(1+\ep)^2} 
- \frac{{\sigma}^{(0)} \otimes \Gamma_2  }{\Gamma(1+\ep)^2} 
- \frac{  \Gamma_1 \otimes {\sigma}^{(0)} \otimes \Gamma_1  }{\Gamma(1+\ep)^2} 
+ \frac{ \Gamma_1 \otimes \sigma^{(1)} }{\Gamma(1+\ep)}   
+ \frac{\sigma^{(1)} \otimes \Gamma_1}{\Gamma(1+\ep)}.    
\end{split}
\label{eq2}
\ee

Although finite, the $\sigma^{(i)}$ still depend on unphysical renormalization and
factorization scales because of the truncation of the perturbative expansion. 
In the following, we will consider for simplicity 
the case of equal renormalization and factorization scales ,
$\mu_r=\mu_f=\mu$. 
The residual $\mu$ dependence is easily determined by solving the 
renormalization group equation order-by-order in $\alpha_s$.
The equation reads
\be
0 = \mu^2\frac {{\rm d}\sigma_{p+p\to H+j}}{{\rm d}\mu^2} = \mu^2\frac{\rm d}{{\rm d}\mu^2} \int {\rm d}x_1 {\rm d}x_2 g(x_1,\mu^2) g(x_2,\mu^2)
\sigma(x_1,x_2,\as(\mu^2),\mu^2).
\ee
The $\mu$-derivative of the right hand side can be computed using the known evolution equations for the strong coupling constant 
and the gluon density
\be
\begin{split} 
& \mu^2\frac{\partial\as}{\partial \mu^2} = -\as\lp \beta_0 \frac{\as}{2\pi} + \beta_1 \lp\frac{\as}{2\pi}\rp^2 + \mathcal O(\as^3)\rp,
\\
& \mu^2\frac{\partial g(\mu^2)}{\partial \mu^2} = 
\frac{\as}{2\pi}\; g(\mu^2)\otimes \lp P_{gg}^{(0)} 
+ \frac{\as}{2\pi} P_{gg}^{(1)} + 
\mathcal O(\as^2) \rp.
\end{split}
\ee
 Solving these renormalization group equations, we get
\be\label{eq:scale}
\begin{split}
\sigma^{(0)}_{\mu_1} &= \sigma^{(0)}_{\mu_2},
\;\;\;\;\;\;\;
\sigma^{(1)}_{\mu_1} = \sigma^{(1)}_{\mu_2} + 
L_{12} \lp
3\beta_0 \sigma^{(0)}_{\mu_2} - P_{gg}^{(0)}\otimes \sigma^{(0)}_{\mu_2} - \sigma^{(0)}_{\mu_2}\otimes P_{gg}^{(0)}\rp, \\
\sigma^{(2)}_{\mu_1} &= \sigma^{(2)}_{\mu_2} 
+ L_{12}\lp
4\beta_0 \sigma_{\mu_2}^{(1)} - P_{gg}^{(0)}\otimes \sigma^{(1)}_{\mu_2} - \sigma^{(1)}_{\mu_2}\otimes P_{gg}^{(0)} + 3 \beta_1 \sigma_{\mu_2}^{(0)} - P_{gg}^{(1)}\otimes \sigma^{(0)}_{\mu_2} + \right. \\
&- \left. \sigma^{(0)}_{\mu_2}\otimes P_{gg}^{(1)} \rp + 
\frac{1}{2} L_{12}^2\lp
12 \beta_0^2 \sigma_{\mu_2}^{(0)} - 7 \beta_0 \lp P_{gg}^{(0)}\otimes \sigma^{(0)}_{\mu_2} + \sigma^{(0)}_{\mu_2}\otimes P_{gg}^{(0)}\rp + \right. \\
&+\left. P_{gg}^{(0)}\otimes P_{gg}^{(0)}\otimes \sigma_{\mu_2}^{(0)} + \sigma_{\mu_2}^{(0)} \otimes P_{gg}^{(0)}\otimes P_{gg}^{(0)}
+  2 P_{gg}^{(0)}\otimes \sigma_{\mu_2}^{(0)} \otimes P_{gg}^{(0)} \rp,
\end{split} 
\ee 
where $\sigma_{\mu}^{(i)} \equiv \sigma^{(i)}(\as(\mu),\mu)$
and $L_{12} = \ln \mu_1^2/\mu_2^2$.

It follows from  Eqs.~(\ref{eq2},\ref{eq:scale}) that, in order to 
obtain $\sigma^{(2)}$ at a generic scale, apart from lower-order results
we need to know the NNLO renormalized cross-section ${\bar \sigma}^{(2)}$ 
and convolutions of NLO and 
LO cross-sections with various splitting functions.
Up to terms induced by the renormalization, 
there are three contributions to  ${\bar \sigma}^{(2)}$ that are  required:
\begin{itemize} 
\item the two-loop virtual corrections to $gg \to Hg$;
\item   the one-loop virtual corrections to 
$gg \to H+gg$; 
\item the double-real contribution  $gg \to H+ggg$.  
\end{itemize} 
We note that helicity amplitudes for 
{\it all} of these processes  are available in the literature. The 
two-loop amplitudes for $gg \to Hg$ were recently 
computed  in 
Ref.~\cite{Gehrmann:2011aa}. 
The one-loop corrections to $gg \to Hgg$~\cite{gghgg_1loop} and the tree amplitudes for $ gg \to Hggg$~\cite{gghggg_tree}
are known. Moreover, in the two latter cases, 
these amplitudes  are available in the form 
of a Fortran code   in the program 
MCFM~\cite{mcfm}.   In principle, they can be just 
taken from MCFM and used with no  
modification in another numerical program.

Since the above discussion implies that {\it all}   ingredients for the 
NNLO computation of $gg \to H+{\rm jet}$ are available and, in fact, have 
been available for some time, it is important 
to understand what has prevented  the community from performing this and 
similar calculations.
In fact, the main
difficulties  with NNLO calculations appear when we attempt 
to combine the different contributions, since integration 
over phase-space introduces 
additional singularities if the required number of jets 
is lower than the parton multiplicity.  
To perform the phase-space  integration, 
we  must  first isolate singularities in tree- and loop amplitudes.  It required a long time to establish a convenient way to do this.

The computational method that we will explain shortly  is based 
on the idea that relevant singularities can be isolated 
using appropriate parametrizations of  phase-space 
and expansions in plus-distributions \cite{Frixione:1995ms,an1}. 
To illustrate this point, we consider 
the integral 
\be
I(\epsilon) = \int \limits_{0}^{1} {\rm d} x x^{-1-a\epsilon} F(x), 
\ee
where the  function $F(x)$ has a well-defined limit
$\lim \limits_{x \to 0}^{} F(x) = F(0)$. We would like to construct the 
Laurent expansion of $I$ in $\epsilon$. This can be accomplished  by 
writing
\be
\frac{1}{x^{1+a\epsilon}} = -\frac{1}{a\epsilon} \delta(x) + \sum_{n=0}^{\infty}
\frac{(-\ep a)^n}{n!}\left[\frac{\ln^n(x)}{x}\right]_+
\ee
so that 
\be
I(\epsilon) = \int \limits_{0}^{1} {\rm d} x \left ( -\frac{F(0)}{a\epsilon}  + \frac{F(x) - F(0)}{x}
 -a\epsilon  \frac{F(x) - F(0)}{x} \ln(x) +...\right ).
\ee
The above equation provides the required Laurent  expansion of the integral $I(\ep)$.  We note that each term in such an expansion can be calculated  independently from 
other terms.   

To use this approach for computing   NNLO QCD corrections, we need 
to map the relevant phase-space  to a unit hypercube 
in such a way that extraction of singularities  is 
straightforward.  It is intuitively clear that correct variables 
to use  are the re-scaled energies of unresolved 
partons and the relative angles between 
two unresolved (collinear) partons. However, 
the problem is that  different partons become unresolved in different parts of the phase-space.  It is not immediately clear how to switch between different 
sets of coordinates and cover the full phase-space.  

We note that for  NLO QCD computations, this  problem was 
solved in  Ref.~\cite{Frixione:1995ms}, where it was 
explained that the  full phase-space can be partitioned into 
sectors in such a way 
that in each sector only one parton ($i$) can  produce 
a soft singularity and only one pair 
of partons ($ij$) can produce a collinear singularity. 
In each sector,  
the proper variables are  the energy of the parton $i$ and the relative 
angle between partons $i$ and $j$. 
Once the partitioning of the phase-space is established and proper variables are chosen for each sector, 
we can use an expansion in plus-distributions to construct  
relevant subtraction terms for each 
sector. With the  subtraction terms in place,  
the Laurent expansion of  cross-sections in  $\epsilon$ can be constructed, 
and each 
term in such an expansion can be integrated 
over the phase-space independently. 
Therefore, partitioning of the 
phase-space into suitable sectors and proper parametrization of the 
phase-space in each of these sectors are the two crucial elements 
needed to extend this method to NNLO.
In the next Section we discuss these issues in detail.

\section{Phase-space parametrizations and sector decomposition} 
\label{phasespace}

\subsection{Phase-space for leading order processes }
\label{phasespace_lo}

We now discuss how to parametrize the leading-order phase-space 
for  the process $g_1 +g_2  \to H + g_3$. 
This will be needed both for the leading-order
cross section and 
for the NLO virtual and NNLO double virtual corrections, so 
it must be computed  in  $d = 4 - 2 \ep $ dimensions.
We note that the integration over the leading order 
phase-space is not singular, because of the requirement that a jet is observed. 
We work in the center-of-mass frame of the two incoming 
gluons, so that their momenta are parametrized as 
\be
p_{1} = \frac{\sqrt{s}}{2} \left ( 1, 0, 0,   1 \right ),
\;\;\;\; 
p_{2} = \frac{\sqrt{s}}{2} \left ( 1, 0, 0, - 1 \right ). 
\ee
The center-of-mass collision energy is denoted by $\sqrt{s}$ and 
the mass of the Higgs boson is denoted by  $m_H$. 
The production cross-section, averaged over spins and colors 
of the two colliding gluons,  is written as 
\be
{\rm d} {\rm \sigma}_{gg \to H+g}  
= \frac{1}{512s}{\rm d}{\rm Lips}_{12 \to H3} |{\cal M}_{gg \to gH}|^2
\times F_j,
\label{eq31}
\ee
where $F_j$ is the ``measurement  function'' that 
restricts the integration to the region of phase-space 
where there is an identified jet.  The amplitude ${\cal M}_{gg \to gH}$ describes 
production of an on-shell Higgs boson in hadronic collisions and 
${\rm d}{\rm Lips}_{12 \to H3}$ is the Lorentz-invariant phase-space. 

The parametrization of the phase-space ${\rm d}{\rm Lips}_{12 \to 3H}$ in Eq.~(\ref{eq31}) is easily obtained 
by integrating over the momentum of the on-shell Higgs boson and 
then over the center-of-mass 
energy of the gluon $g_3$. We find 
\be
{\rm d}{\rm Lips}_{12 \to 3H}
= \frac{{\rm d} \Omega^{(d-2)}_3  p_{\perp,H}^{-2\ep}\;{\rm d} \cos \theta_3 }{8 (2\pi)^{d-2}}
\left ( 1-\frac{m_H^2}{s} \right ),\;\;\;\;\;
p_{\perp,H} = E_{\rm max} \sin \theta_3,
\label{phspLO}
\ee
where $E_{\rm max} = (s-m_H^2)/(2\sqrt{s})$.
With this parametrization, the momentum of the gluon $g_3$ reads 
\be
p_3 = E_{\rm max}  \left ( 1 , \vec n_3 \right ), 
\label{genpar}
\ee
where 
$n_3 = \left ( \sin \theta_3 \cos \varphi_3, 
\sin \theta_3 \sin \varphi_3, \cos \theta_3 \right)$,
and the Higgs boson momentum is obtained from 
momentum conservation: $p_H = p_1 + p_2 - p_3$.
Note that in Eq.~(\ref{phspLO}),  $p_{\perp,H}$ 
is the transverse momentum of the Higgs boson relative to the collision 
axis. 

Before proceeding further, we note that  the azimuthal 
angle $\varphi_3$ of the emitted gluon is a dummy variable, 
since neither matrix element  squared  nor the measurement function 
$F_J$ depend on it for our choice of $p_{1}$ and $p_2$.  
Hence, we can rotate it away by taking the $g_3$ momentum to be 
\be
p_3 = E_{\rm max}  \left ( 1, \sin \theta_3,
0,\cos \theta_3 \right),
\ee
and integrate over ${\rm d} \Omega^{(d-2)}_3$  
in Eq.~(\ref{phspLO}). Once this is done, we can set this solid angle 
to its four-dimensional expression to simplify calculations at 
higher orders.  This is legitimate to do as long as 
we can identify this angle, associated with global rotations of  final 
states in the plane transverse to the collision 
axis,  when parametrizing higher-multiplicity phase-spaces.  Nevertheless, to maintain sufficiently general leading order kinematics, we 
re-introduce the azimuthal angle $\varphi_3$ and keep it to 
generate momenta of the gluon and the Higgs boson. This amounts 
to  writing 
${\rm d} \Omega^{(d-2)}_3 \to {\rm d} \varphi_3$ 
in Eq.~(\ref{phspLO}) and then using Eq.~(\ref{genpar}) for 
the gluon $g_3$ momentum.  In addition, 
since the transverse momentum of the Higgs boson is an 
observable quantity, the differential cross-section 
${\rm d}\sigma/{\rm d}p_{\perp,H}$ should 
be finite for each value of $p_{\perp,H}$ to all orders 
in perturbation theory.   Hence, we can divide the cross-section 
 by $p_{\perp,H}^{-2\ep}$ without changing the final result.  
This amounts to removing this factor from the phase-space parametrization 
at both leading and higher orders in perturbation theory.  Putting all these remarks together, we conclude that we can 
choose the leading order phase-space to be ``four-dimensional,''
\be
{\rm d}{\rm Lips}_{12 \to 3H} \to 
\frac{{\rm d} \cos \theta_3 {\rm d} \varphi_3 }{32\pi^2}
\left ( 1-\frac{m_H^2}{s} \right ) =  \left ( 1-\frac{m_H^2}{s} \right ) 
\frac{{\rm d} x_3 {\rm d} x_4}{8\pi},
\ee
where we 
introduced $\cos \theta_3 = 1- 2x_3$ and $\varphi_3 = 2 \pi x_4$ to parametrize 
the momentum of gluon $g_3$ as given by Eq.~(\ref{genpar}).
We must remember to normalize NLO and NNLO phase-spaces to 
$p_{\perp,H}^{-2\ep}$ for consistency. 

Although we will not discuss this in any detail in this paper, we note that 
it is straightforward to include decays of the Higgs boson. Indeed, because 
the Higgs boson   momentum is an observable quantity, all singularities should cancel out in 
the differential cross-section ${\rm d} \sigma/{\rm d} \vec p_H$. 
Once this differential cross-section is known 
and because the Higgs boson is a scalar particle, so that no spin correlations are 
present,  we can easily turn ${\rm d} \sigma/{\rm d} \vec p_H$ into quantities such as
${\rm d}\sigma/{\rm d} \vec p_{\gamma_1}{\rm d} \vec p_{\gamma_2}$ by letting the Higgs boson 
decay in its rest frame and then boosting the four-momenta of the two photons 
into the center-of-mass frame.  

\subsection{Phase-space for next-to-leading order processes }
\label{phasespace_nlo}

In this Section, we 
consider the parametrization of the phase-space for the process 
$g_1 + g_2 \to H + 
g_3 + g_4$.  This process represents a real-emission 
contribution  to the production cross-section of  the Higgs and 
one jet at next-to-leading order. It is also important for the NNLO 
computation where  integration of 
one-loop corrections to $ gg \to Hgg$ amplitudes over  
the $ggH$ phase-space is required.
  
As we already explained  in the Introduction, a good  parametrization 
of the phase-space ${\rm dLips}_{g_1g_2 \to H g_3 g_4}$ should facilitate 
the extraction of singularities  from the matrix elements of the 
process $g_1g_2 \to H g_3g_4$.  
Of particular importance are {\it collinear} singularities.
We can compare two cases: {\it i})  $g_4$ is emitted collinear  to $g_1$; 
{\it ii}) $g_4$ is emitted  collinear to $g_3$. In the first 
case, it is easiest to extract the singularity  if the $z$-axis 
is chosen to coincide with the direction of the gluon $g_1$ and in the second 
case with the direction of the gluon $g_3$.  This immediately tells us that 
a suitable parametrization of the phase-space should  depend on the 
 kinematics of the process.  As we mentioned  in the Introduction, 
this is  the main 
idea behind the FKS subtraction method~\cite{Frixione:1995ms}. 

Following Ref.~\cite{Frixione:1995ms}, we note that the first 
step towards a convenient phase-space pa\-ra\-metrization is the 
phase-space partitioning.  The goal of such a partitioning 
is to create sectors where one and only one  gluon  or 
one and only one pair of gluons 
can become unresolved.  Once we know which gluon or which pair of gluons
can produce singularities,  we choose the energy of the 
potentially soft gluon and the relative angle between the two potentially 
collinear gluons as the primary variables for the phase-space parametrization 
in the given sector.  To illustrate this procedure, we begin by removing  the
symmetry between the two gluons 
in the final state by separating them into ``resolved''  
and ``unresolved'' ones.  To this end, we introduce the following function 
of transverse momenta of gluons $g_3$ and $g_4$, 
\be
\Delta^{(i)}_{p_{\perp}} = \frac{p_{\perp,j}}{ p_{\perp,3} + p_{\perp,4}},
\;\;\;j \ne i,
\ee
and write
\be
\frac{1}{2!} {\rm d}{\rm Lips}_{12 \to 34H}  =
\frac{1}{2!} {\rm d}{\rm Lips}_{12 \to 34H} \left (
\Delta_{p_\perp}^{(4)} + \Delta_{p_\perp}^{(3)} \right ) 
\to  {\rm d} {\rm Lips}_{12 \to 34H}
\Delta_{p_\perp}^{(4)}. 
\label{as_psp}
\ee
In the last step we used the fact that the phase space, the kinematic 
constraints on final-state particles and all matrix elements 
are symmetric 
with respect to permutations of gluons $g_3$ and $g_4$.
Given the structure of the damping 
factor $\Delta_{p_\perp}^{(4)}$,  it is clear 
that singularities of the matrix element related 
to gluon $g_3$  are unimportant, 
and we only need to  consider cases 
when gluon $g_4$ becomes either soft {\it or} 
collinear  to one of the three hard directions defined  by the momenta $g_1,g_2$ and $g_3$.  Note that $g_3$ and $g_4$ cannot 
both be soft, or collinear to the collision axis at the same time, because 
we require a jet in the final state. 
To separate the  collinear-singular regions, we introduce another 
partition  of unity
\be
1 = \Delta_{\theta}^{(41)} + \Delta_{\theta}^{(42)} + \Delta_{\theta}^{(43)}.
\label{partnlo}
\ee
In Eq.~(\ref{partnlo}), we use 
\be
\Delta_{\theta}^{(4i)} = \frac{\rho^{j4} \rho^{k4}}{ 
\rho^{14} \rho^{24} + \rho^{14}\rho^{34} + \rho^{24} \rho^{34} 
},\;\;\;\;\;j,k \ne i,4,
\label{partnlo1}
\ee
where $\rho^{ij} = 1- \vec n_i \cdot \vec n_j $ and 
$\vec n_i$ is the three-vector that parametrizes momentum direction 
of the particle $i$.  Again, the $\Delta_{\theta}^{(4i)}$ are labeled in such a way 
that the subscript indicates a pair of particles that can become collinear
without forcing the angular damping factor to vanish.  Inserting this partition of unity Eq.~(\ref{partnlo})
into the phase-space of Eq.~(\ref{as_psp}), we 
obtain 
\be
\frac{1}{2!} {\rm d}{\rm Lips}_{12 \to 34H}  
\to  \sum \limits_{i=1}^{3} 
{\rm d} {\rm Lips}_{12 \to 34H}^{(4i)},
\;\;\;\;
{\rm d} {\rm Lips}_{12 \to 34H}^{(4i)} = 
{\rm d} {\rm Lips}_{12 \to 34H}
\Delta_{p_\perp}^{(4)} \Delta_{\theta}^{(4i)}.
\ee
The above decomposition defines pre-sectors that we will 
refer to as ${\rm Sc}^{(4i)}$. A phase-space parametrization for each 
of these pre-sectors is chosen in such a way that the soft and collinear 
singularities that are relevant for that pre-sector 
can be extracted in the easiest  possible way.

We now describe these parametrizations explicitly. 
In general, we will parametrize the phase-spaces by splitting them 
into  ``regular'' and ``singular'' parts
\be
{\rm d} {\rm Lips}_{12 \to 34H}^{(4i)} = 
\Delta_{p_\perp}^{(4)} \Delta_{\theta}^{(4i)} {\rm d}{\rm Lips}_{Q(12)  \to 3H}
 \times [{\rm d} g_4]^{(4i)}.
\ee
The regular NLO phase-space is the same for all 
pre-sectors.  It includes all particles except the (potentially soft) 
gluon $g_4$. We write it as 
\be
{\rm d}{\rm Lips}_{Q(12)  \to 3H} 
= \frac{{\rm d} x_4 {\rm d} x_5}{(8\pi)}
\frac{2 E_{g_3}}{ (Q_0  - \vec Q \cdot \vec n_3 )}
\left ( 
\frac{E_{g_3}^2 \sin^2 \theta_3}{p_{\perp,H}^2} \right )^{-\ep}, 
\label{reg_phsp_nlo}
\ee
where we have introduced the notation   $Q = p_1 + p_2 - p_4$ and 
$p_3 = E_{g_3} ( 1, \vec n_3) $. Also, 
\be
\begin{split} 
& E_{g_3} = \frac{Q^2 -m_H^2}{2(Q_0 - \vec Q \cdot \vec n_3 )},\;\;\; 
 {\vec n}_3 = \left ( \sin \theta_3 \cos \varphi_3, \sin \theta_3 \sin \varphi_3, 
\cos \theta_3  \right ), \\
& \cos \theta_3 = 1 - 2 x_4,\;\;\;
\sin \theta_3 = +\sqrt{1 - \cos^2 \theta_3},\;\;\; 
\varphi_3 = 2 \pi x_5.
\end{split} 
\ee
Following the discussion of the leading order phase-space parametrization, 
we have dropped the $\epsilon$-dependent 
part of the integral over azimuthal angle of the gluon $g_3$, and have
normalized the remaining $\epsilon$-dependent 
part of the phase-space to the transverse momentum 
of the Higgs boson.

Parametrization of the singular phase-space depends on the 
pre-sector.  To explain this, we begin by  considering  
pre-sector ${\rm Sc}^{(41)}$.
To parametrize the singular phase-space
for this pre-sector, we note that, thanks to the damping 
factors,   the singularities occur when $g_4$ is collinear  to $g_1$ 
or when $g_4$ is soft.  Hence, it is convenient to choose the parametrization 
where the energy of the gluon $g_4$ and the relative angle between the 
three-momenta of $g_4$ and $g_1$ are  basic variables. The azimuthal 
angle of the gluon $g_4$ is conveniently defined relative to the plane 
formed by the $g_1$ and $g_3$ three-momenta.  We therefore write 
\be
p_4 = E_{g_4} \left ( 1, \sin \theta_4 \cos \varphi_4 ,
\sin \theta_4 \sin \varphi_4 , \cos \theta_4 \right ),
\ee
where $ \varphi_4 = {\tilde \varphi}_4 + \varphi_3 $. The singular 
phase-space reads 
\be
[{\rm d} g_4]^{(41)} = \frac{E_{g_4}^{d-3}{\rm d}E_{g_4} 
{\rm d} \cos \theta_4 {\rm d}\varphi_4 (\sin^2 \theta_4 \sin^2 {\tilde \varphi}_4)^{-\ep}
{\rm d} \Omega^{(d-3)}}{2 (2\pi)^{d-1}}.
\ee
The $(d-3)$-dimensional solid angle does not enter any of the scalar products 
and therefore can be integrated away.  We write 
\be
E_{g_4} = E_{\rm max} x_1,\;\;\; 
\cos \theta_{4} = 1 - 2 x_2,
\label{eq3.10}
\ee
where $E_{\rm max}$ is introduced after Eq.~(\ref{phspLO}).
The singular phase-space for ${\rm Sc}^{(41)}$  becomes 
\be
[{\rm d} g_4]^{(41)} = 
E_{\rm max}^{d-2} \; \frac{2^{-2\ep} \Omega_{4}^{(d-2)}}{(2 \pi)^{d-1} }
\;\;
x_1^{1-2\ep} x_2^{-\ep} (1-x_2)^{-\ep} 
\frac{{\rm d} \varphi_4 (\sin^2 (  \varphi_4 - \varphi_3) )^{-\ep}}
{\int \limits_{0}^{2\pi} {\rm d} 
\varphi_4 (\sin^2 ( \varphi_4))^{-\ep}}.
\ee
We use 
\be
\int \limits_{0}^{2\pi} {\rm d} \varphi_4 (\sin^2 \varphi_4)^{-\ep}
= 2^{1-2\ep} B\left ( \frac{1}{2}-\ep,\frac{1}{2}-\ep \right ),
\ee
and  write $\varphi_4 = 2 \pi x_3$ to find 
\be
\begin{split}
& [{\rm d} g_4]^{(41)} = \left ( 1- \frac{\pi^2}{3}\ep^2 
- 2\zeta_3 \ep^3 + \frac{\pi^4}{90}\ep^4
\right )
\frac{\Gamma(1+\ep)}{(4\pi)^{d/2}}\; 2^{-2\ep}
\left (  2 E_{\rm max} \right ) ^{2-2\ep} \;\\
& \times 
 x_1^{1-2\ep} x_2^{-\ep} (1-x_2)^{-\ep} (\sin^2 ( \varphi_4 - \varphi_3))^{-\ep} 
\prod \limits_{1}^{3} {\rm d} x_i . 
\end{split} 
\ee
Combining everything, we find the expression for the phase-space 
of the pre-sector ${\rm Sc}^{(41)}$ to be 
\be
{\rm d} {\rm Lips}_{12 \to 34H}^{(41)} 
= {\rm Norm} \times {\rm PS}_w  \times
{\rm PS}^{-\ep} 
\frac{{\rm d} x_1 {\rm d} x_2 {\rm d} x_3
{\rm d} x_4 {\rm d} x_5
}{
x_1^{1+2\ep} x_2^{1+\ep}
}
\times 
\left [ x_1^2 x_2 \right ], 
\label{eq_sp_41}
\ee
where  $ {\rm PS} = 16 E_{\rm max}^2 E_3^2 
\sin^2 \theta_3(1 - x_2)(\sin^2 ( \varphi_4 - \varphi_3)) 
/ p_{\perp,h}^2 $ and 
\be
\label{nlo_norm}
\begin{split} 
& {\rm Norm} = \frac{\Gamma(1+\ep)}{(4\pi)^{d/2}}
\left ( 1- \frac{\pi^2}{3}\ep^2 
- 2\zeta_3 \ep^3 + \frac{\pi^4}{90}\ep^4
\right ), 
\;\;\;\;
\\
& {\rm PS}_w = 
\frac{E_{\rm max}^2 E_3}{8 \pi^2
\left ( \sqrt{s} - E_4 ( 1 - \vec n_3 \cdot \vec n_4  ) \right )}
\Delta_{\theta}^{(41)} \Delta_{p_\perp }^{(4)}.
\end{split} 
\ee

These equations  allow us to generate four-momenta  of all final-state particles. 
Indeed, a set of random numbers $x_1,..,x_5$ 
gives us momenta of the gluon $g_4$ and the direction 
of the unit vector $\vec n_3$ 
that parametrizes the momentum direction 
  of a  ``hard'' gluon  $g_3$ in the center-of-mass frame
of colliding gluons.  
Using this information,  
we can find the energy of the gluon $g_3$ and  determine the momentum of the 
Higgs boson  from momentum conservation.

The phase-space of the second pre-sector ${\rm Sc}^{(42)}$ 
${\rm d} {\rm Lips}_{12 \to 34H}^{(42)}$ is parametrized in a similar
 way, except that we now need a simple parametrization of the relative angle 
between gluons $g_2$ and $g_4$. Therefore, we write 
$\cos \theta_4 = -1 + 2 x_4$. This is the only change that occurs at the 
level of momentum generation and everything else, including 
the phase-space parametrization, can be borrowed from 
Eq.~(\ref{eq_sp_41}).

The phase-space parametrization for  the third pre-sector  
${\rm Sc}^{(43)}$ requires some  changes.  The main difference with respect to the previous cases is that 
now the collinear  direction corresponds to the ``hard'' final state gluon $g_3$, which 
means that we need to choose the relative angle between $g_3$ and $g_4$ as the primary 
variable for phase-space parametrization. 
In the reference frame where the momentum of gluon $g_3$ is 
along the $z$-axis, the direction of the gluon $g_4$ is chosen to be
\be
\vec n_{4,3||z} = 
\left ( \sin \theta_4 \cos \varphi_4, \sin \theta_4 \sin \varphi_4, 
\cos \theta_4 \right ).
\label{eq3.19}
\ee
The phase-space parametrization employs angles $\theta_4$ and $\varphi_4$. 
The momentum of the 
gluon $g_4$ 
 in the center-of-mass reference frame is obtained by rotating Eq.~(\ref{eq3.19}) 
in the $x-z$ plane by $\theta_3$ and in the $x-y$ plane by $\varphi_3$. 
We parametrize the energy of the gluon $g_4$ and its relative 
angle with respect to  $g_3$ using Eq.~(\ref{eq3.10}).  We conclude that 
 the parametrization of  ${\rm Sc}^{(43)}$  phase-space coincides 
with Eq.~(\ref{eq_sp_41}) except that in ${\rm PS}^{-\ep}$, we should substitute 
$ (\sin^2 (\varphi_4 - \varphi_3 ))^{-\ep} \to 
(\sin^2 (\varphi_4))^{-\ep}.$

The above formulae can be used to construct phase-space 
parametrizations for next-to-leading 
computations or for the calculation  of the one-loop corrections to 
$gg \to H+gg$ process.  In the latter case, 
one should be careful since 
it is customary for one-loop virtual corrections 
to be normalized with  the factor 
\be
c_\Gamma = \frac{\Gamma(1+\ep) \Gamma(1-\ep)^2}{(4\pi)^{2-\ep} \Gamma(1-2\ep)}.
\label{cgamma}
\ee
If we choose  the  normalization in such a way that 
one power  of $\Gamma(1+\ep)/(4\pi)^{d/2}$ is factored out
{\it per loop}, the expression for ${\rm Norm}$ in 
Eq.~(\ref{nlo_norm}) changes.  To use Eq.~(\ref{eq_sp_41}) for 
the computation of real-virtual corrections, we should 
make the following replacement there
\be
{\rm Norm} \to 
{\rm Norm}_{\rm RV} \equiv c_\Gamma {\rm Norm}  = \frac{\Gamma^2(1+\ep)}{(4\pi)^d}
\left ( 1 - \frac{\pi^2}{2}\ep^2 - 4\zeta_3 \ep^3 + 
\frac{\pi^4}{24} \ep^4
\right ).
\ee

\subsection{Phase-space for next-to-next-to-leading order processes }

In this Section we consider the partonic  process $g_1+g_2 
\to H +  g_3 + g_4 + g_5$ 
and discuss how to generate the  phase-space in a way 
that facilitates the extraction of singularities.  
We begin with a discussion of the phase-space 
partitioning.  Similar to the one-loop case, we first 
partition  the phase-space in a way that allows us to identify the 
``hard''  gluon by writing
\be
\Delta_{p_\perp}^{(ij)} = \frac{p_{\perp,k}}{p_{\perp,3}+p_{\perp,4} + p_{\perp,5}},\;\;\;
i \ne j \ne k,\;\;\; i,j,k \in [3,4,5].
\ee
Because  $\Delta^{(34)}+ \Delta^{(35)} + \Delta^{(45)} = 1$,  
we can use this partition  of 
unity and the symmetry of  the phase-space, the measurement functions 
and  the matrix elements with respect to 
permutations of gluons $g_{3},g_{4}$ and $g_{5}$, to write 
\be
\begin{split}
& \frac{1}{3!} {\rm d}{\rm Lips}_{12 \to H345 }
 = \frac{1}{3!} {\rm d}{\rm Lips}_{12 \to H345 }
\left ( \Delta_{p_\perp}^{(34)} + \Delta_{p_\perp}^{(35)} + \Delta_{p_\perp}^{(45)} \right ) 
= \frac{1}{2!} {\rm d}{\rm Lips}_{12 \to H345 } \Delta_{p_\perp}^{(45)}
\\
& = {\rm d}{\rm Lips}_{12 \to H345 } \Delta_{p_\perp}^{(45)} \theta(E_{g_4} - E_{g_5}).
\label{eq3:30}
\end{split}
\ee
In the last step we introduced the energy ordering of the two gluons; 
this allows us to 
remove the final symmetry factor.
 
We must next partition the phase-space  to extract collinear singularities. 
To do so, we closely follow the discussion of the next-to-leading order 
case in the previous Section. We split the phase-space 
into nine different sectors that we denote by the  possible collinear directions of the gluons $4$ and $5$.
We have three triple-collinear sectors $4||5||i$, with $i = 1,2,3$ and 
six double-collinear sectors $4||i \otimes 5||j$, where $i \ne j \in [1,2,3]$.
To write the weight for each of the nine sectors, we introduce the auxiliary quantities
\be
\begin{split}
& d_{i \in [4,5]} = \sum \limits_{j=1}^{3} \rho_{ij},\;\;\; 
d_{i \in [4,5]k} = \sum \limits_{j=1,j\ne k}^{3} \rho_{ij},\;\;\; 
d_{45ij} = \rho_{45} + \rho_{4i} + \rho_{5j}.
\end{split}
\ee
Denoting  the weight of a sector where gluon $4$ is allowed to become collinear to  gluon $i$ and 
gluon $5$ to gluon $j$ by $w_{4i;5j}$,
we write ($ k \ne n \ne 4 \ne 5 \ne i \ne j$)
\be
\begin{split}  
w_{4i;5j}|_{i = j} & = \frac{\rho_{4k}\rho_{4n} \rho_{5k} \rho_{5n}}{d_4 d_5} \Bigg [ 
         \left ( \frac{1}{d_{4k}}+\frac{1}{d_{4n}} \right )
          \left( \frac{1}{d_{5k}}+\frac{1}{d_{5n}} \right )             
\\
&   
        + \left (\frac{1}{d_{4i}}+\frac{1}{d_{4k}} \right )\left (\frac{1}{d_{5k}}
             +\frac{1}{d_{5n}} \right ) \frac{\rho_{4i}}{d_{45ni}} 
        +\left (\frac{1}{d_{4i}}+\frac{1}{d_{4n}} \right )\left (\frac{1}{d_{5k}}
      +\frac{1}{d_{5n}} \right ) \frac{\rho_{4i}}{d_{45ki}}   
\\
&
        +\left (\frac{1}{d_{4k}}+\frac{1}{d_{4n}} \right ) 
      \left ( \frac{1}{d_{5i}}+\frac{1}{d_{5k}}  \right ) \frac{\rho_{5i}}{d_{45in}}  
      + \left ( \frac{1}{d_{4k}}+\frac{1}{d_{4n}} \right ) \left ( \frac{1}{d_{5i}}
        +\frac{1}{d_{5n}} \right ) \frac{\rho_{5i}}{d_{45ik}} \Bigg ],
\end{split}
\ee
and ($ k \ne n \ne 4 \ne 5 \ne i$, $l \ne m \ne 4 \ne 5 \ne j$)
\be\label{dc_weight}
 w_{4i;5j}|_{i \ne j} = 
\frac{\rho_{4k} \rho_{4n} \rho_{5l} \rho_{5m} }{d_4 d_5} 
\left (\frac{1}{d_{4k}}+\frac{1}{d_{4n}} \right ) \left (
  \frac{1}{d_{5l}}+\frac{1}{d_{5m}} \right ) \frac{\rho_{45}}{d_{45ij}}.
\ee
Using Eq.~(\ref{eq3:30}), we decompose the phase-space 
as 
\be
 \frac{1}{3!} {\rm d}{\rm Lips}_{12 \to H345 }
= \sum \limits_{\alpha \in S}^{} {\rm d}{\rm Lips}_{12 \to H345 }^{(\alpha)},
\ee     
where $S = [(41;51),(42;52),(43;53),(41;52),(42;51),(41;53),(43;51),
(42;53),(43;52)]$ and 
\be
{\rm d}{\rm Lips}_{12 \to H345 }^{(\alpha)}
= 
{\rm d}{\rm Lips}_{12 \to H345 } 
\; \Delta_{p_\perp}^{(45)} \; \theta(E_{g_4} - E_{g_5})
\;w_{\alpha}.
\ee

We now discuss the parametrization of the phase-spaces for 
individual pre-sectors.  Because of the $\Delta_{p_\perp}^{(45)}$ 
factor, we consider  gluon $g_3$ as part of the regular phase-space
and gluons $g_4,g_5$ as part of the singular phase-space. 
Regular phase-spaces are the same for all pre-sectors and are 
parametrized  in the same way as at NLO in Eq.~(\ref{reg_phsp_nlo}), 
except that the vector  $Q$ in that equation 
becomes $Q = p_1 + p_2 - p_4 - p_5$. 

We begin with the triple-collinear sectors.  We have three such sectors ${\rm Sc}^{(4i;5i)}$, $i \in [1,2,3]$.  
In these sectors, 
singularities can appear if gluons $g_{4,5}$ are soft, and 
if they are  collinear to the 
direction $\vec n_i$,  or to each other.  The phase-space 
parametrization should enable us to extract all of these 
singularities.  We will start the discussion with 
the triple-collinear initial sector ${\rm Sc}^{(41;51)}$.  

The first step is to find 
 independent degrees 
of freedom, which is non-trivial because we have to perform 
computations in dimensional regularization.  To illustrate this 
point, we use   $d$-dimensional rotational invariance to   choose  
the momenta  of five gluons as follows
\be
\begin{split}
& p_{1,2} = \frac{\sqrt{s}}{2} \left ( 1, 0, 0, \pm 1;0 \right ),
\\
& 
p_3 = E_{g_3} \left ( 1, \sin \theta_3 \cos {\tilde \varphi}_3, 
\sin \theta_3 \sin {\tilde \varphi}_3, 
\cos \theta_3; 0 \right ),
\\
& p_4 = E_{g_4} \left (1, \sin \theta_4, 0, \cos \theta_4; 0 \right ),
\\
& p_5 = E_{g_5} \left ( 1, \sin \theta_5 \cos \varphi_5, \sin \theta_5 \sin \varphi_5
\cos \alpha, 
\cos \theta_5; \sin \theta_5 \sin \varphi_5 \sin \alpha \right ).
\label{eq_ep1}
\end{split}
\ee
Note that these momenta 
are shown as {\it five-dimensional} vectors; the fifth component 
corresponds to one of the axes in the $(d-4)$-dimensional space. The 
angle $\alpha$ parametrizes leakage into the $(d-4)$-dimensional vector space. 
Note also that we have chosen to give the $(d-4)$-dimensional 
component to the {\it softer} of the two gluons. The reason 
for this choice will be explained shortly.  With this 
parametrization, the angular part of the phase-space 
becomes
\be
\begin{split}
 {\rm d} \Omega_{g_3}^{(d-1)} 
{\rm d} \Omega_{g_4}^{(d-1)} 
{\rm d} \Omega_{g_5}^{(d-1)}  
& \sim 
{\rm d}[\cos \theta_3] (\sin^2 \theta_3)^{-\ep}
 {\rm d} \tilde \varphi_3 (\sin^2 {\tilde \varphi}_3)^{-\ep}
{\rm d}\Omega_{g_3}^{(d-1)} \;
\\
& 
\times {\rm d}[ \cos \theta_4] ( \sin^2 \theta_4)^{-\ep}
{\rm d}\Omega_{g_4}^{(d-2)} \; 
{\rm d}[ \cos \theta_5] (\sin ^2 \theta_5)^{-\ep}
\\
& \times {\rm d} \varphi_5 (\sin^2 \varphi_5)^{-\ep} 
{\rm d} [\cos \alpha] ( \sin^2 \alpha)^{-1-\ep}
{\rm d}\Omega_{g_4}^{(d-4)}. 
\end{split}
\label{eq_ep_2}
\ee

We can generalize the momentum parametrization in Eq.~(\ref{eq_ep1}) 
by rotating all momenta in the $xy$-plane by the angle $\varphi_4$. 
Obviously,  the momenta of the incoming gluons 
$p_{1,2}$ do not change, while the other momenta become
\be
\begin{split}
& p_4 = E_4 \left (1, \sin \theta_4\cos \varphi_4, 
\sin \theta_4 \cos \varphi_4, \cos \theta_4; 0 \right ),
\\
& p_5 = E_5 \left ( 1, \sin \theta_5 \cos_\alpha(\varphi_4+\varphi_5), 
\sin \theta_5 \sin_\alpha(\varphi_4 + \varphi_5), 
\cos \theta_5; \sin \theta_5 \sin \varphi_5 \sin \alpha \right ),
\\
& 
p_3 = E_3 \left ( 1, \sin \theta_3 \cos \varphi_3
\sin \theta_3 \sin \varphi_3, 
\cos \theta_3; 0 \right ).
\label{eq_ep_3}
\end{split}
\ee
In Eq.~(\ref{eq_ep_3}),  we have introduced the notation 
\be
\begin{split}
& \cos_\alpha(\varphi_4 + \varphi_5) = 
\cos \varphi_4 \cos \varphi_5 
- \sin \varphi_4 \sin \varphi_5 \cos \alpha, 
\\
& \sin_\alpha(\varphi_4 + \varphi_5) = 
\sin \varphi_4 \cos \varphi_5  + \cos \varphi_4 \sin \varphi_5
\cos \alpha,
\end{split} 
\ee
and $\varphi_3 = {\tilde \varphi}_3 + \varphi_4$.  Note that the phase-space 
is written in terms of ${\tilde \varphi}_3$, the relative azimuthal 
angle  of $g_4$ and $g_3$, and that 
\be
\cos_{\alpha}^2(a) + \sin_\alpha^2(a) \ne 1.
\ee

Before we express the phase-space parametrization 
in terms of suitable variables, we make a few general comments.
We note that our choice of the phase-space 
parametrization and assignment of extra-dimensional 
components is restricted by two requirements:
\begin{itemize} 
\item extra-dimensional 
components and angles should not complicate the extraction of 
singular limits;
\item  extra-dimensional momenta components should 
not appear in the non-singular matrix elements and kinematic constraints.
\end{itemize} 
It turns out that the parametrization of the momenta in Eq.~(\ref{eq_ep_3}) 
satisfies the first requirement for the triple-collinear sector 
${\rm Sc}^{(41;51)}$.  This happens because the parametrization is chosen 
in such a way that the 
scalar products $p_1 \cdot p_4, p_1 \cdot p_5, p_4 \cdot p_5$ 
that can potentially lead to singularities in this sector 
do not depend on the 
extra-dimensional angle $\alpha$.

We now discuss how to satisfy the second requirement. We note that 
full parametrization of Eq.~(\ref{eq_ep_3}) is not needed for the
highest multiplicity $gg\to Hggg$ hard matrix element. Indeed, 
a configuration where all the three final-state gluons are resolved is
non-singular, hence we can use a $d=4$ phase-space parametrization to describe it. We will
see explicitly below that this amounts to setting 
 $\alpha=0$ in Eq.~(\ref{eq_ep_3}).
Therefore, we only have to 
explain how to satisfy the second requirement in configurations 
where one or both of $g_4$, $g_5$ are unresolved.  To this end, we note that  in all soft limits 
this requirement is automatically satisfied. 
Indeed, since $E_{g_4} \to 0$ 
implies $E_{g_5} \to 0$, in  any of the soft limits the gluon 
momentum with the $\ep$-dimensional component 
is not present in the hard matrix element and in kinematic 
constraints.   The $\alpha$-dependence 
will therefore reside solely in the unresolved phase-space and 
in eikonal factors and splitting functions. 
It is important that this dependence on $\alpha$ is non-singular, so that 
the numerical integration can be performed in a  straightforward way.

The collinear limits are more complicated. If $p_5$ is collinear to either 
$p_1$ or $p_4$, then $\varphi_5 = 0$ or $\theta_5 = 0$, which implies 
that the $\ep$-dimensional components of momenta and the 
dependence on $\alpha$  disappear 
from the matrix elements. On the other hand, this does not 
mean that collinear limits are independent of $\alpha$. 
Indeed, such a
dependence is present  in the spin-correlation 
part of the splitting functions.  We must account for that in the computation. 
This can be done in a straightforward way 
since this dependence is non-singular.  
Finally, consider the kinematic situation where 
$p_4$ is collinear to $p_1$ and 
$p_5$ is {\it resolved }. In this case,
 the matrix element  squared becomes 
\be
|{\cal M}|^2 \approx \frac{1}{p_1\cdot p_4} (2 C_A g_s^2) P_{gg}^{\mu \nu}(p_4,\kappa_4) {\cal M}^{\mu}(p_{14},p_2,p_3,p_5) 
{\cal M}^{*,\nu}(p_{14},p_2,p_3,p_5),
\label{eq_prob}
\ee
where $p_{14} = p_1 - p_4$ and 
$\kappa_4$ is the 
spin-correlation vector that tells us how the collinear direction is approached 
(see Section~\ref{singularlimits} or~\cite{Catani:1996vz} for details).
Eq.~(\ref{eq_prob})  
implies that the matrix element 
depends on the four-vector $p_5$ and, according to Eq.~(\ref{eq_ep_3}), 
$p_5$   has $\ep$-dimensional components. This dependence is 
unfortunate, since it becomes   unclear how to use four-dimensional 
methods, such as spinor-helicity techniques, to simplify 
calculations of scattering amplitudes in that situation.   
However, when $p_1||p_4$ we are left with only three different
directions $n_1$, $n_3$, $n_5$.  We can use $d-$dimensional rotational 
invariance
to remove any 
$\ep$-dimensional components from the matrix elements 
in Eq.~(\ref{eq_prob}).    To do so, we first remove 
the $y$-component of $p_3$ by 
rotating  all momenta in the $xy$-plane by the angle 
$-\varphi_3$.  
This rotation does not change $p_{14} \sim (1,0,0,1)$ and 
$p_2 \sim (1,0,0,-1)$.  
We then perform another rotation in the $y\ep$-plane, 
to remove the $\ep$-dependent component of the vector $p_5$. Because 
none of the momenta  in the matrix element has both $y$- and 
$\ep$-dimensional components, such a rotation 
does not change $p_{14},p_2$ and $p_3$, while 
it makes $p_5$ four-dimensional. 
We note that, although we rotated away the 
$\ep$-dimensional components of the resolved four-vectors  that are  used 
in the hard matrix elements,  these vectors still depend on the 
$\ep$-dimensional angle  $\alpha$.  
In addition,  because of spin correlations, we 
also must rotate the vector $\kappa_4^{\mu} 
= (0, \cos \varphi_4, \sin \varphi_4, 0,0)$ 
that enters $P_{gg}^{\mu \nu}$ in Eq.~(\ref{eq_prob}).  This rotated vector 
receives $\ep$-dimensional components and becomes $\alpha$-dependent. 
The purpose of the rotation therefore is to move the $\ep$-dimensional 
components from the resolved momenta in the matrix element to 
the splitting function,  where it is easy to account for them explicitly.
Finally, we stress  that the 
very possibility to rotate away the $\ep$-dimensional components 
of particle momenta is connected to the rotational invariance of spin-summed 
scattering 
amplitudes squared in $d$-dimensional space-time.  This seems to suggest that the 
easiest framework in which to implement this techniques is conventional dimensional
regularization, where the momenta of all external particles {\it and} 
their polarization vectors are treated as $d$-dimensional.  We will
discuss this point in more detail shortly.

We now discuss the explicit parametrizations of the relevant phase-spaces. 
For the sector ${\rm Sc}^{(41;51)}$, the  singular phase-space reads 
\be
\begin{split} 
& [{\rm d} g_4] [{\rm d}g_5] \theta(E_{g_4} - E_{g_5}) 
= 
\frac{{\rm d} \Omega^{(d-3)} {\rm d} \Omega^{(d-4)}}{2^{4+2\ep} (2\pi)^{2d-2}}
{\rm d} \varphi_4 \left [ \sin^2(\varphi_4 - \varphi_3) \right ]^{-\ep} 
{\rm d} \cos \alpha \left [ \sin^2 \alpha \right ]^{-1-\ep}
\\
&  
\times 
\left [ \xi_1 \xi_2  \right ]^{1-2\ep}
\left [\eta_4 ( 1 - \eta_4 ) \right ]^{-\ep} 
\left [\eta_5 ( 1 - \eta_5 ) \right ]^{-\ep} 
\left [ \lambda (1 - \lambda ) \right ]^{-1/2-\ep}
\frac{|\eta_4 - \eta_5|^{1-2\ep}}{D^{1-2\ep}}
\\
& \times
\left ( 2 E_{\rm max} \right )^{4-4\ep} \theta(\xi_1 - \xi_2)
\theta\left (\xi_{\rm max}  - \xi_2  \right ) 
\; {\rm d} \xi_1 {\rm d} \xi_2 {\rm d} \eta_4 {\rm d} \eta_5
{\rm d} \lambda.
\end{split} 
\label{eq_337}
\ee
The variables 
introduced in the above formula parametrize the energies and angles of 
the (potentially) unresolved gluons in the following way
\be
E_{g_4,g_5} = E_{\rm max} \xi_{1,2},
\;\;\;\;\;\;\;
\xi_{\rm max} = {\rm min} 
\left [ 1, \frac{1-\xi_1}{1- (1 - m_h^2/s) \xi_1 \eta_{45}} \right ],
\ee
and 
\be
\begin{split} 
& \eta_{45} = \frac{|\eta_4 - \eta_5|^2}{D},\;\;\;\;\;
\sin^2 \varphi_{5} = 4 \lambda (1-\lambda) \frac{|\eta_4 - \eta_5|^2}{D^2},
\\
& D = \eta_4 + \eta_5 - 2\eta_4 \eta_5 + 2 ( 2\lambda - 1) \sqrt{\eta_4 \eta_5 (1-\eta_4) (1-\eta_5)}.
\end{split} 
\ee
The two variables $\eta_{4,5}$ are scalar products of the reference direction 
vector $\vec n_{1}$ and the vectors that parametrize directions 
of the two gluons
\be
2 \eta_{4,5} = 1 - \vec n_{4,5} \cdot \vec n_{1}.
\ee
The parametrization of triple-collinear phase-spaces in Eq.~(\ref{eq_337}) 
is still too complicated 
to extract all  singularities; further decomposition is required.  This is 
achieved by a sequence 
of variable changes  that we describe below, 
following Refs.~\cite{Czakon:2010td,Czakon:2011ve}.  Specifically, 
we split the triple-collinear initial-initial sector into five sectors
\be
{\rm d Lips}^{(41;51)} = \sum \limits_{i}^{5} {\rm dLips}^{(41;51,i)}.
\ee
To project onto individual 
contributions, we need to perform the following changes 
of variables 
\be
\begin{split} 
& {\rm Sc}^{(41;51,1)}:\;\; \xi_1 = x_1,\; \xi_2 = x_1 x_{\rm max} x_2,\; 
\eta_4 = x_3,\;\; \eta_5 = \frac{x_3 x_4}{2},
\\
& {\rm Sc}^{(41;51,2)}:\;\; \xi_1 = x_1,\; \xi_2 = x_1 x_{\rm max} x_2,\; 
\eta_4 = x_3,\;\; \eta_5 = x_3 \left ( 1- \frac{x_4}{2} \right ),
\\
& {\rm Sc}^{(41;51,3)}:\;\; \xi_1 = x_1,\; \xi_2 = x_1 x_{\rm max} x_2 x_4,\; 
\eta_4 = \frac{x_3 x_4}{2} ,\;\; \eta_5 = x_3, \\
& {\rm Sc}^{(41;51,4)}:\;\; \xi_1 = x_1,\; \xi_2 = x_1 x_{\rm max} x_2,\;
\eta_4 = \frac{x_3 x_4 x_2}{2},\;\; \eta_5 = x_3, \\
& {\rm Sc}^{(41;51,5)}:\;\;  \xi_1 = x_1,\; \xi_2 = x_1 x_{\rm max} x_2,\; 
\eta_4 = x_3 \left (1 - \frac{x_4}{2} \right ) ,\;\; 
\eta_5 = x_3.
\end{split} 
\ee
We also write $\lambda = \sin^2(\pi x_5/2)$. 
This change of variables introduces  
a factor of $\pi$ in the
normalization of the phase-space that is included in the 
expressions below.  The integration region for $x_5$  is always 
between zero and one.

We also note that the $(d-4)$-dimensional angle $\alpha$ introduces 
singularities 
in the phase-space parametrization. 
To take care of them, we calculate the 
integral over this  
angle,
\be
I_\alpha 
= \int \limits_{-1}^{1} \frac{{\rm d} \cos \alpha}{[\sin^2 \alpha]^{-1+\ep}}
 = \frac{1}{2^{1+2\ep}} \int \limits_{0}^{1} \frac{{\rm d} x_9 }{
x_9^{1+\ep} (1-x_9)^{1+\ep}}
= \frac{\Gamma(-\ep)^2}{2^{1+2\ep} \Gamma(-2\ep)},
\ee
and write 
\be
\begin{split} 
& \frac{{\rm d} \left [ \cos \alpha \right ]}{[\sin^2 \alpha]^{1+\ep}} 
= I_\alpha \times \frac{\Gamma(1-2\ep)}{2 \Gamma(1-\ep)^2} (-\ep) 
\frac{{\rm d} x_9}
{x_9^{1+\ep} (1-x_9)^{1+\ep}}
 \\
& \to 
I_\alpha 
\times \frac{\Gamma(1-2\ep)}{ \Gamma(1-\ep)^2} (-\ep) 
\frac{{\rm d} x_9 (1-x_9)^{-\ep} }
{x_9^{1+\ep}},
\end{split}
\label{eq_x11}
\ee
where $\cos \alpha = 1-2 x_9$ and in the last step we used the symmetry 
of the matrix element with respect to
$x_9 \leftrightarrow 1-x_9$, to simplify the integrand. 
We can expand Eq.~(\ref{eq_x11}) 
in plus-distributions.  Such an expansion 
does not introduce additional poles in $\ep$.  We find 
\be
-\frac{\ep}{x_9^{1+\ep}}  = 
\delta(x_9) - \ep \left [ \frac{1}{x_9} \right ]_+ + .. 
\ee
Note that the first term in the expansion corresponds to 
$\alpha = 0$, which reduces the 
parametrization of momenta of all  final-state particles to their 
four-dimensional limits.  
The ``extra-dimensional'' 
momenta components 
and the ``extra-dimensional'' angles 
appear with an additional suppression in $\ep$, but because 
of infra-red singularities, they start contributing 
to differential cross-sections 
already  at ${\cal O}(\ep^{-2})$. 

For each of the five sectors ${\rm Sc}^{(41;51,i)}$, we write the phase-space 
in the form 
\be
{\rm dLips}_{41;51}^{(i)} \sim {\rm Norm} \times {\rm  PS}_{w,i} {\rm PS}_{i}^{-\ep} 
\times \frac{(-\ep)}{x_9^{1+\ep}}\prod \limits_{k=5}^{9} {\rm d}x_k
\times \prod \limits_{j=1}^{4}  \frac{{\rm d} x_j}{x_j^{1+a_{j}^{(i)} \ep}}
\times \left [ x_1^{b_1^{(i)}} x_2^{b_2^{(i)}} x_3^{b_3^{(i)}} x_4^{b_4^{(i)}} \right ].
\ee
Below we present the functions 
${\rm PS}_{w,i}$, ${\rm PS}_i$ and the exponents $a_{j=1...4}^{(i)}$ and 
$b_{j=1...4}^{(i)}$ for each of the sectors. First, 
we note that the normalization factor is common to  all sectors; it reads 
\be
{\rm Norm} = \left [ \frac{\Gamma(1+\ep)}{(4\pi)^{d/2}} \right ]^2 
\left ( 1 - \frac{\pi^2}{2} \ep^2 - 2\zeta(3) \ep^3 + \frac{3 \pi^4}{40} \ep^4 
\right ).
\ee
We also note that we can write 
\be
\begin{split} 
& {\rm PS}_{w,i} = \frac{1}{2\pi^2} 
\frac{E_3 E_{\rm max}^4}{Q_0 - \vec Q \cdot \vec n_3}\; {\overline {\rm PS}}_{w,i},\;\;\;
{\rm PS}_i = 
\frac{1024 E_3^2 \sin^2 \theta_3 E_{\rm max}^4 (1-x_9)}{\mu^4 p_{\perp, H}^2} 
\sin^2\left ( \varphi_{43} \right ) \; {\overline {\rm PS}_i},
\label{eq352}
\end{split} 
\ee
where $\varphi_{43} = \varphi_4 - \varphi_3$.  The expressions for the exponents and the phase-space factors for each of the 
five  sectors read (we suppress the sector label everywhere in the equations 
below)
\be
\begin{split}
 {\rm \underline{Sector}~}{\rm Sc}^{(41;51,1)}:& \;\; 
\{ a_1 = 4, a_2 = 2, a_3 = 2, a_4 = 1 \}, \;\;\;\;
\{b_1 = 4, b_2 = 2, b_3 = 2, b_4 = 1\}; \\
& {\overline {\rm PS}}_w = \frac{(1-\frac{x_4}{2})x_{\rm max}^2}{2N_1(x_3,\frac{x_4}{2},\lambda)
},\\
& {\overline {\rm PS}} = \frac{x_{\rm max}^2
\left (1-\frac{x_3x_4}{2}  \right )
\lambda (1 - \lambda) \left ( 1- \frac{x_4}{2} \right )^2
(1-x_3)
}{2 N_1^2(x_3,\frac{x_4}{2},\lambda)}.
\end{split} \nonumber
\ee

\be
\begin{split}
  {\rm \underline{Sector}~}{\rm Sc}^{(41;51,2)}&:\;\;
 \{ a_1 = 4, a_2 = 2, a_3 = 2, a_4 = 2 \}, \;\;\;\;
\{b_1 = 4, b_2 = 2, b_3 = 2, b_4 = 2\}; \\
& {\overline {\rm PS}}_w = \frac{x_{\rm max}^2}{4
N_1(x_3,1-\frac{x_4}{2},\lambda)},
\;\;\;
\\
& {\overline {\rm PS}} = 
\frac{x_{\rm max}^2  (1-x_3) \left (1-\frac{x_4}{2}  \right )
\left ( 1-x_3(1-\frac{x_4}{2}) \right )
\lambda (1 - \lambda)}{4 N_1^2(x_3,1-\frac{x_4}{2},\lambda)}. 
\end{split} \nonumber
\ee

\be
\begin{split}
 {\rm \underline{Sector}~}{\rm Sc}^{(41;51,3)} & :\;\;
 \{ a_1 = 4, a_2 = 2, a_3 = 2, a_4 = 3 \}, \;\;\;\;
\{b_1 = 4, b_2 = 2, b_3 = 2, b_4 = 3\}; \\
& {\overline {\rm PS}}_w = 
\frac{x_{\rm max}^2(1-\frac{x_4}{2})}{2 N_1(x_3,\frac{x_4}{2},\lambda)},
\\
& {\overline {\rm PS}} = \frac{x_{\rm max}^2
(1-x_3) \left (1 - \frac{x_3 x_4}{2} \right ) 
\left (1-\frac{x_4}{2}  \right )^2
\lambda (1 - \lambda)
}{2 N_1^2(x_3,\frac{x_4}{2},\lambda)}.
\end{split} \nonumber
\ee

\be
\begin{split}
 {\rm \underline{Sector}~}{\rm Sc}^{41;51,4} & :\;\;
 \{ a_1 = 4, a_2 = 3, a_3 = 2, a_4 = 1 \}, \;\;\;\;
\{b_1 = 4, b_2 = 3, b_3 = 2, b_4 = 1\}; \\
& {\overline {\rm PS}}_w = 
\frac{x_{\rm max}^2(1-\frac{x_2x_4}{2} )}{2
N_1(x_3,\frac{x_4 x_2}{2},\lambda)},
\\
& {\overline {\rm PS}} = \frac{x_{\rm max}^2
(1-\frac{x_2x_3x_4}{2}) \left (1 - \frac{x_2 x_4}{2} \right )^2 
\left (1-x_3  \right )
\lambda (1 - \lambda)
}{2 N_1^2(x_3,\frac{x_2 x_4}{2},\lambda)}.
\end{split} \nonumber
\ee

\be
\begin{split}
 {\rm \underline{Sector}~}{\rm Sc}^{(41;51,5)}&:\;\;
 \{ a_1 = 4, a_2 = 2, a_3 = 2, a_4 = 2 \}, \;\;\;\;
\{b_1 = 4, b_2 = 2, b_3 = 2, b_4 = 2\}; \\
& {\overline {\rm PS}}_w = 
\frac{x_{\rm max}^2}{4
N_1(x_3,1-\frac{x_4}{2},\lambda)},
\\
& {\overline {\rm PS}} = \frac{x_{\rm max}^2
(1-x_3) ( 1-\frac{x_4}{2} ) ( 1-x_3(1-\frac{x_4}{2} ) )
\lambda (1 - \lambda)
}{4 N_1^2(x_3,1-\frac{x_4}{2},\lambda)}.
\end{split}
\ee
The function $N_1$ reads 
\be
        N_1(x_3,x_4,\lambda) = 1 + x_4 (1 -2 x_3)  
        - 2(1-2 \lambda)\sqrt{x_4(1-x_3)(1-x_3 x_4)}.
\ee
The above phase-space parametrization is such that the limits 
$x_{i} \to 0$, $i = 1...4$ of the matrix element squared can be 
easily computed; we will discuss this in more detail in the next Section. 
In the remainder of this Section, we will focus on the phase-space 
parametrization of the other  pre-sectors.  

We note that the phase-space parametrization for the  
triple-collinear  pre-sector ${\rm Sc}^{42;52}$ 
is constructed in exact analogy to ${\rm Sc}^{41,51}$. The only difference 
is that the collinear direction is now $\vec n_2 = (0,0,-1)$ instead of 
$\vec n_1 = (0,0,1)$. This means that, in terms of the $\eta$-variables, angles 
of gluons $g_{4,5}$ relative to the collision axis are given by $\cos \theta_{4,5} = 
-1 + 2\eta_{4,5}$. 

The construction of the 
phase-space parametrization for the triple-collinear pre-sector 
${\rm Sc}^{(43;53)}$   is
slightly more involved, since the collinear direction now is the direction 
of the gluon $g_3$. It is therefore convenient to write momenta of $g_4$ and $g_5$ 
in the reference frame where $g_3$ is along the $z$-axis. We write 
\be
\begin{split} 
& p_{3}^{(z)} = E_{g_3} \left ( 1,0,0,1 ; 0 \right ) 
\\
& p_{4}^{(z)} = E_{g_4} \left ( 1, \sin \theta_4 \cos \varphi_4, \sin \theta_4 \sin \varphi_4 , 
\cos \theta_4 ; 0 \right),
\\
&
p_5^{(z)} = E_{g_5} \left ( 1, 
\sin \theta_5 \cos_{\alpha} ( \varphi_{45} ), 
\sin \theta_5 \sin_{\alpha} (\varphi_{45} ), 
\cos \theta_5 ; \sin \theta_5 \sin \varphi_5   \sin \alpha
\right ),
\end{split}
\label{eq350}
\ee 
where $\varphi_{45} = \varphi_4 + \varphi_{5}$. In this sector, the  
scalar products whose vanishing leads to singularities 
are $p_3 \cdot p_4, p_3 \cdot p_5$ and 
$p_4 \cdot p_5$. It is easy to see from Eq.~(\ref{eq350}) that 
these scalar products are independent of $\alpha$.  The phase-space for ${\rm Sc}^{43;53}$ 
depends on two relative  angles $\varphi_4$ and $\varphi_5$, 
so that ${\rm Lips}_{43;53} \sim \left ( \sin^2 \varphi_4 \sin^2 \varphi_5 \right )^{-\ep}$.
To get the momenta in  the center-of-mass frame, we rotate these 
vectors first in the $xz$ plane 
by $\theta_3$, and then in the $xy$ plane by $\varphi_3$.   
The parametrization of the singular 
phase-space is similar to what we have 
discussed in connection with ${\rm Sc}^{(41;51)}$, except that 
the collinear direction now is $\vec n_3$.

We finally turn to the discussion of the double-collinear sectors. 
First, consider the sectors 
where collinear singularities arise from emission along two 
incoming particles, (${\rm Sc}^{(41;52)}$ and ${\rm Sc}^{(42;51)}$).   
In such sectors, scalar products whose vanishing may create singularities 
are $p_{4,5} \cdot p_1 $ and $p_{4,5} \cdot p_2$.  Vanishing 
of the scalar product 
$p_4 \cdot p_5$ cannot lead to singularities  in this sector, see Eq.~(\ref{dc_weight}).  With this in mind, 
we parametrize  momenta of the three final-state 
gluons in the center-of-mass frame as 
\be
\begin{split} 
& p_3 = E_{g_3} \left ( 1, \sin \theta_3 \cos \varphi_3, \sin \theta_3 \sin \varphi_3, \cos \theta_3; 0 \right ),
\\
& p_4 = E_{g_4} \left ( 1, 
\sin \theta_4 \cos (\varphi_3 + {\tilde \varphi}_4), 
\sin \theta_4 \sin (\varphi_3 + {\tilde \varphi}_4), 
\cos \theta_4 ; 0  \right ), 
\\
& p_5 = E_{g_5} \left ( 1, 
\sin \theta_5 \cos_{\alpha} (\varphi_3 + {\tilde \varphi}_5), 
\sin \theta_4 \sin_{\alpha} (\varphi_3 + {\tilde \varphi}_5 ), 
\cos \theta_5 ;   \sin \theta_5 \sin \varphi_5 
\sin \alpha \right ). 
\end{split}
\ee
The phase-space is parametrized  in terms of 
the relative angles ${\tilde \varphi}_{4}$ and 
${\tilde \varphi}_5$. We find 
\be
\begin{split} 
& [{\rm d} g_4 ] [{\rm d} g_5] \theta(E_{g_4} - E_{g_5})
 = \frac{E_{\rm max}^{2d-4}}{4(2\pi)^{2d-2}} \Omega_4^{(d-3)} \Omega_5^{(d-3)}
 \theta(\xi_1 - \xi_2) 
\theta(\xi_{\rm max} - \xi_2 ) 
\\
& 
\times {\rm d}\xi_1 {\rm d} \xi_2  \xi_1^{1-2\ep} \xi_2^{1-2\ep} 
{\rm d} \cos \theta_4 {\rm d} \cos \theta_5  ( \sin^2 \theta_4)^{-\ep} 
( \sin^2 \theta_5 ) ^{-\ep}
\\
& \times 
\frac{{\rm d} \varphi_4 (\sin^2 \left ( {\tilde \varphi_4} \right ) )^{-\ep}}{
\int \limits_{0}^{2\pi} {\rm d} {\tilde \varphi_4} (\sin^2 \varphi_4)^{-\ep}}
\frac{{\rm d} {\tilde \varphi_5} (\sin^2 {\tilde \varphi_5})^{-\ep}}{
\int \limits_{0}^{2\pi} {\rm d} \varphi_5 (\sin^2 \varphi_5)^{-\ep}
}
\times 
\frac{{\rm d} \left [ \cos \alpha \right ]}{[\sin^2 \alpha]^{1+\ep}}, 
 \end{split} 
\ee
where $E_{g_4,g_5} = E_{\rm max} \xi_{1,2}$.
We now change variables $\xi_1 = x_1, \xi_2 = x_1 x_2 x_{\rm max}$, 
$\cos \theta_{4,5} = 1 - 2 x_{3,4}$,  ${\tilde \varphi_{4,5}} = 2 \pi x_{5,6}$ 
and $\cos \alpha = 1 -2 x_9$. We use  symmetry with respect to 
$x_9 \to 1-x_9$ to simplify  
the expression for the phase-space. We obtain 
\be
{\rm dLips}_{41;52}  \sim {\rm Norm} \times {\rm  PS}_{w} {\rm PS}^{-\ep} 
\times \frac{(-\ep)}{x_9^{1+\ep}}\prod \limits_{k=5}^{9} {\rm d}x_k
\times \prod \limits_{j=1}^{4}  \frac{{\rm d} x_j}{x_j^{1+a_{j}\ep}}
\times 
\left [ 
x_1^{b_1} x_2^{b_2} x_3^{b_3} x_4^{b_4} 
\right ].
\ee
The normalization factors read\footnote{We note that 
in Ref.~\cite{Czakon:2011ve} the double collinear sectors are further 
split by an additional partitioning of energy and angle variables. We 
find that such a partitioning is unnecessary.}
\be
\begin{split} 
& {\rm Norm} = \left [ \frac{\Gamma(1+\ep)}{(4\pi)^{d/2}} \right ]^2 
\left ( 1 - \frac{\pi^2}{2} \ep^2 - 2\zeta(3) \ep^3 + \frac{3 \pi^4}{40} \ep^4 
\right ),
\\
& {\rm PS}_w = \frac{1}{2\pi^2} \frac{E_3 E_{\rm max}^4}{Q_0 - \vec Q \cdot \vec n_3}\; {\overline {\rm PS}}_w,
\\
& {\rm PS} = \frac{2^{8} 
E_3^2 \sin^2 \theta_3 E_{\rm max}^4 x_{\rm max}^2 }{\mu^4 p_{\perp,  H}^2} 
\sin^2 {\tilde \varphi}_4 \sin^2 {\tilde \varphi}_5  ( 1- x_3) ( 1 - x_4)(1-x_9),
\end{split} 
\label{eq359}
\ee
and the exponents read
\be
 \{ a_1 = 4, a_2 = 2, a_3 = 1, a_4 = 1 \}, \;\;\;\;
\{b_1 = 4, b_2 = 2, b_3 = 1, b_4 = 1\}.
\ee

The other type of double-collinear sectors that need to be considered is the initial-final one. We focus
for definiteness on ${\rm Sc}^{(41;53)}$.  The momenta
read 
\be
\begin{split} 
& p_3 = E_{g_3} \left ( 1, \sin \theta_3 \cos \varphi_3, 
\sin \theta_3 \sin \varphi_3, \cos \theta_3; 0 \right ),
\\
& p_4 = E_{g_4}  \left ( 1, \sin \theta_4 \cos (\varphi_3 + {\tilde \varphi_4}), 
 \sin \theta_4 \sin (\varphi_3 + {\tilde \varphi_4} ). 
\cos \theta_4 \right ), 
\\
& p_5^{(z)}  = E_{g_5} \left ( 1, \sin \theta_5 \cos {\tilde \varphi_5}, \sin \theta_5 \sin {\tilde \varphi_5}  
\cos \alpha, 
\cos \theta_5; 
\sin \theta_5 \sin {\tilde \varphi_5} \sin \alpha \right ). 
\end{split}
\ee
Note that $p_{3}$ and $p_4$ are given in the center-of-mass frame, 
while $p_5$ is  written in the reference frame where $p_3$ 
is along the $z$-axis.  To obtain $p_5$  in the center-of-mass 
frame, we rotate it by $\theta_3$ in the $xz$-plane and by $\varphi_3$ 
in the $xy$-plane.  The phase-space is identical to Eq.~(\ref{eq359}).
The discussion of all other double collinear sectors proceeds along the same lines.

\section{Singular limits}
\label{singularlimits}

In this Section, we describe the extraction 
of singular limits.
We begin with the next-to-leading order computation. We note that 
we will {\it not} discuss the most general case from the point of view 
of color correlations; instead 
we will make use of the fact that we are studying Higgs boson 
production in association with a jet and  so the number of colored 
particles never exceeds five.  
This feature leads to simplification of the color correlations 
in soft  limits.  We will make use of these simplifications 
in what follows. 

\subsection{Limits at next-to-leading order}
\label{lim_nlo}

Consider, for definiteness,  the NLO sector ${\rm Sc}^{(43)}$.
 The phase-space for this sector,  ${\rm dLips}^{(43)}_{12 \to 34H}$,
is given by an expression  similar 
to Eq.~(\ref{eq_sp_41}), where $x_2$ parametrizes the relative angle between 
$g_4$ and $g_3$. We have to integrate 
the matrix element squared $|{\cal M}_{gg \to Hgg}|^2$ over the phase-space. 
The integration has the form
\be
\int \limits_{0}^{1} \frac{{\rm d} x_1}{x_1^{1+2 \ep}}\frac{{\rm d} x_2}{x_2^{1+\ep}}
\;... \times F(x_1,x_2,... ), \;\;\;\;
F(x_1,x_2,...)  = \left [ x_1^2 x_2 \right ] |{\cal M}_{gg \to Hgg}|^2,
\label{eq_345}
\ee
where the ellipses denote the measurement 
 function, regular parts of the phase-space,
various damping factors and possible additional arguments of the 
function $F$.  All of these things are not important for discussing the 
structure of singularities which is shown  explicitly in Eq.~(\ref{eq_345}). The singularities 
correspond to $x_1=0$ or $x_2=0$, and the function $F(x_1,x_2,..)$ 
is  finite in those limits.  The integral in Eq.~(\ref{eq_345}) is calculated using an expansion in plus-distributions, 
as we explained in Section~\ref{setup}.
It follows that in order to perform the integration in Eq.~(\ref{eq_345}), 
we need to understand 
values of the function  $F(x_1,x_2,..)$ in cases when one (or both) of the two first 
arguments  vanishes.

Consider first the $x_1 = 0$ limit. According to the phase-space 
parametrization described in Section~\ref{phasespace_nlo}, 
$x_1 = 0$ implies that $g_4$ is soft: $E_{g_4} = 0$. 
In the soft limit, the matrix element is written as 
a product of a reduced matrix element and the eikonal factor
\be
|{\cal M}_{g_1g_2\to H g_3g_4} |^2
\approx 2 C_A g_s^2   \left ( 
{\cal I}_{12;4}^{(0)} + {\cal I}_{13;4}^{(0)} 
+ {\cal I}_{23;4}^{(0)}
 \right ) |{\cal M}_{g_1g_2\to H g_3} |^2, 
\label{eq_start00}
\ee
where 
\be
{\cal I}^{(0)}_{ij;k} = 
S_{ij}(p_k) = \frac{p_i \cdot p_j}{(p_i \cdot p_k) (p_j \cdot p_k)},
\label{eq_ijk}
\ee
is the eikonal factor. To calculate  the soft $x_1 \to 0$ limit, 
we note that the eikonal factor is quadratic in 
$p_4 = E_{g_4}(1,\vec n_4) \sim x_1$ 
and so it is easy to compute the required limit. We obtain 
\be
F(0,x_2,...) = \frac{C_A g_s^2}{E_{\rm max}^2} 
\left ( 
\frac{\rho_{12} \rho_{34} }{\rho_{14} \rho_{24}}
+
\frac{\rho_{13} }{\rho_{14}} 
+
\frac{\rho_{23} }{\rho_{24}} 
\right )|{\cal M}_{gg \to Hg_3}|^2,
\label{limnlosoft}
\ee 
where we traded $x_2$ for $\rho_{34}/2$ which is valid in 
sector ${\rm Sc}^{(43)}$. We note that 
potential singularities that correspond to gluon $g_4$  being 
collinear to gluons $g_1$ or $g_2$
are apparent in Eq.~(\ref{limnlosoft}); these singularities are 
removed by the angular 
damping factor Eq.~(\ref{partnlo1}) for this sector.  

The second singular limit 
we have to consider is  $x_2 =0$. In sector  ${\rm Sc}^{(43)}$,
$x_2 = 0$ means that 
gluon $g_4$ is collinear to gluon $g_3$.  The corresponding 
limit reads 
\be
|{\cal M}_{g g\to H g g} |^2
\approx
\frac{2 C_A g_s^2}{p_3\cdot p_4}  P^{(gg)}_{\mu \nu}(z,\ep) {\cal M}_{gg \to Hg}^{\mu} 
{\cal M}_{gg \to Hg}^{*,\nu}, 
\label{limnlocol}
\ee
where $z = E_{g_4}/(E_{g_3} + E_{g_4})$ and 
\be\label{Pgg0munu}
P^{(gg)}_{\mu \nu}(z,\kappa_4,\ep) = 
-g_{\mu \nu} \left ( \frac{z}{1-z} + \frac{1-z}{z} \right ) 
+ 2 (1-\ep) z (1-z) \kappa_{4,\mu} \kappa_{4,\nu} 
\ee
is the gluon splitting function. The vector $\kappa_{4,\mu}$ 
is the normalized remnant of the momentum 
$p_4$ that parametrizes the projection of $p_4$ on 
the plane transverse to the collinear direction  which in this case 
is fixed to be the momentum of gluon $g_3$. Because of the 
chosen parametrization of $p_4^{\mu}$ at next-to-leading order, 
$\kappa_4^\mu$ has only 
four-dimensional components. 

We will now show how to simplify  Eq.~(\ref{limnlocol}). 
The idea is to trade the 
sum over the Lorentz indices $\mu$ and $\nu$ for a sum over helicity indices. 
This is achieved by inserting the completeness relation 
\be
\sum \epsilon_\lambda^\mu \epsilon_{\lambda}^{\mu '} = 
- g_d^{\mu \mu'}  + \frac{
p_3^{\mu} {\tilde n}^{\mu'} 
+ p_3^{\mu'} {\tilde n}^{\mu}}{p_3 \cdot {\tilde n}},
\ee
where $g_d^{\mu \nu}$ denotes the metric tensor 
of the $d$-dimensional vector space and ${\tilde n}$ is an auxiliary 
vector such that $p_3 \cdot {\tilde n} \ne 0$.
Next, we write 
\be
\begin{split} 
P^{(gg)}_{\mu \nu}(z,\ep) {\cal M}_{gg \to Hg}^{\mu} {\cal M}_{gg \to Hg}^{*,\nu} & 
= - P^{(gg)}_{\mu\nu} \left ( 
\sum \epsilon_\lambda^\mu \epsilon_{\lambda}^{\mu '} - 
\frac{p_3^{\mu} \tilde n^{\mu'} + p_3^{\mu'} \tilde n^{\mu}}{p_3 \cdot \tilde n} \right )
{\cal M}^{\mu'} {\cal M}^{*,\nu} 
\\
& 
= - P^{(gg)}_{\mu\nu} 
\sum \epsilon_\lambda^\mu \epsilon_{\lambda}^{\mu '} 
{\cal M}^{\mu'} {\cal M}^{*,\nu},  
\end{split} 
\ee
where the last step follows from the transversality of 
the physical amplitude  ${\cal M}^{\mu} p_{3,\mu} = 0$ and from $\kappa_{4} 
\cdot p_3 = 0$. Repeating the same procedure with 
the index $\nu$, we find 
\be
P^{(gg)}_{\mu \nu}(z,\ep) {\cal M}^{\mu} {\cal M}^{*,\nu}  
= \sum \limits_{\lambda,\lambda'}
P^{(gg)}_{\lambda \lambda'}(z,\ep) {\cal M}_{\lambda } {\cal M}_{\lambda'}^{*},  
\ee
where the sum over  physical helicities in $d$-dimensional 
space-time is performed. 

We now explain how to compute $P^{(gg)}_{\lambda \lambda'}$. 
First, we note that the polarization vectors of a gluon with 
four-dimensional momenta embedded in a $d$-dimensional space-time  
can be chosen in the following way.  We take the polarization 
vectors to be either four-dimensional vectors that describe 
states of plus and minus helicity, or $(d-4)$ dimensional vectors of the 
type $\epsilon^\mu = (0,0,0,0;0,..,1,0,..,0)$, where projection on a single 
extra-dimensional direction is non-vanishing. 
The helicity-dependent 
splitting function $P^{(gg)}_{\lambda \lambda'}(z,\ep)$ reads 
\be
P^{(gg)}_{\lambda \lambda'}(z,\ep) = 
 \delta_{\lambda \lambda'} \left ( \frac{z}{1-z} + \frac{1-z}{z} \right ) 
+ 2 (1-\ep)z (1-z) 
{\left ( \epsilon_\lambda \cdot \kappa_4 \right )  
\left ( \epsilon^*_{\lambda'} \cdot \kappa_4  \right )}.
\label{pgg_ll}
\ee
For regular $\pm$ polarizations, both terms in Eq.~(\ref{pgg_ll}) are
in general
non-vanishing. For extra-dimensional polarizations, 
$\epsilon_\lambda \cdot \kappa_4  = 0$, because $\kappa_4$ in this 
case is a four-dimensional vector,  and 
$P^{(gg)}_{\lambda \lambda'}  \sim \delta_{\lambda \lambda'}$ for these polarizations. 

We can now compute $P^{(gg)}(z,\ep)$ and calculate  
the $x_2 = 0$ limit of the  function $F_2(x_1,x_2)$. The final result reads 
\be
F_2(x_1,0,...) = 
\frac{2 C_A  g_s^2  x_1}{E_3 E_{\rm max}} 
\sum \limits_{\lambda,\lambda'}
P_{\lambda \lambda'}^{(gg)}(z,\ep) 
{\cal M}^{\lambda}_{gg \to {\tilde g}_3 H} 
{\cal M}^{\lambda'}_{gg \to {\tilde g}_3 H}, 
\label{col_limit}
\ee
where ${\tilde g}_3$ means that the matrix element should be 
computed with the momentum 
of the final state gluon given by the sum of the momenta of the gluons $g_3$ and $g_4$.   We note that Eq.~(\ref{col_limit}) requires the computation 
of scattering amplitudes  for the $gg \to Hg$ process when  $\lambda$ parametrizes an extra-dimensional polarization vector. 
We explain how to do this in Section~\ref{ampl_ep}.

Finally, we discuss how the vector $\kappa_4$ 
is computed. This vector  
parametrizes how the collinear limit 
is approached  in the plane transverse to the collinear direction.
For this reason, it depends on the considered sector. 
To make this explicit, we consider the sector ${\rm Sc}^{(43)}$ and write  
$p_4 = x p_3 + y {\tilde p}_3 + k_\perp \kappa_4$, 
where ${\tilde p}_3 = (E_3, -\vec p_3)$, 
 $\kappa_4 \cdot p_3 = 0$ and  $\kappa_4 \cdot {\tilde p}_3 = 0$. 
A simple computation gives 
$
\vec \kappa_4 = (\cos \theta_3 \cos \varphi_3 \cos \varphi_4 - \sin \varphi_3 \sin \varphi_4, 
\cos \theta_3 \sin \varphi_3 \cos \varphi_4 + \cos \varphi_3 \sin \varphi_4, - \sin \theta_3 \cos \varphi_4)$.
The analogous vectors for the other sectors are much simpler. For 
example, for sectors ${\rm Sc}^{(41)}$ and ${\rm Sc}^{(42)}$, 
we find $\vec \kappa_4 = \left ( \cos \varphi_4, \sin \varphi_4,0 \right  )$.  
We note that these vectors are 
uniquely determined for each of the phase-space points; this allows us to 
construct the correct splitting function and 
perform the {\it local subtraction} of singularities.
The quality of the subtraction terms so constructed  will be studied in Section~7.

These are the only two limits that are required for a NLO computation.  An expression for $F(0,0,...)$ can be easily obtained from the soft limit 
Eq.~(\ref{limnlosoft}), which is non-singular for $\eta_{34}\to 0$. 
Note that the collinear limit has a well-known $1/(1-z)$ singularity as
gluon $g_4$ becomes soft; therefore, to compute $F(0,0,...)$ from the collinear 
limit, one has to cancel $x_1$ in the numerator in Eq.(\ref{col_limit}) 
with $1/(1- z) \sim 1/x_1$ in the splitting function.

\subsection{Limits of double-real emission processes}
\label{lim_nnlo}

In this Section, we briefly discuss the singular 
limits of the double-real emission processes.
As already pointed out, the phase-space partitioning 
splits the phase-space into   
double-singular and triple-singular sectors.  Collinear 
singularities of the double-collinear 
sectors are given by products of gluon splitting functions, 
because the two unresolved gluons 
must be collinear to different directions. 
On the contrary, in the triple-collinear sectors, 
the $1 \to 3$ gluon splitting functions~\cite{Catani:1999ss} are 
required to  describe collinear limits. 
For both double-singular and triple-singular 
sectors, soft singularities  originate from 
both double-soft and single-soft limits. 

We begin by discussing 
the double-soft limit 
of the $g_1 g_2 \to H g_3 g_4 g_5$ scattering amplitude. It occurs 
when the momenta of $g_4$ and $g_5$ become vanishingly small. 
In general, 
double-soft limits 
involve 
color-correlated matrix elements, 
but  in our case this does not occur.  The reason 
is that, once gluons 
$g_4$ and $g_5$ decouple,  the matrix element depends on  
three colored particles, $g_1,g_2,g_3$.  If, following 
Ref.~\cite{Catani:1996vz},  we 
denote the color charge of a particle $i$ by the operator 
$\vec T_i$,  color conservation  implies 
\be
\vec T_{g_1}  + \vec T_{g_2} + \vec T_{g_3} = 0.
\ee
In addition, the squares of the color charge operators are equal to the Casimir 
operators  of the $SU(3)$ gauge group. For gluons, this means 
$\vec T_{g_i}^2 = C_A$. Using these two equations, we 
find 
\be
\vec T_{g_1} \cdot \vec T_{g_2} 
=
\vec T_{g_1} \cdot \vec T_{g_3} 
=
\vec T_{g_2} \cdot \vec T_{g_3} = -C_A/2,
\ee
so that all color correlations are absent. 
As a result, we can use a  simple formula 
for the double soft limit ($S_p = [12,13,23]$)
\be
\begin{split} 
& |{\cal M}_{g_1 g_2 \to H g_3 g_4 g_5}|^2 
\approx  C_A^2 g_s^4 
 \Bigg [
\left ( \sum \limits_{ij \in S_p}^{} S_{ij}(p_4) \right ) 
\left ( \sum \limits_{kn \in S_p}^{} S_{kn}(p_5) \right ) 
\\
&
+ 
\sum \limits_{ij \in S_p}^{} S_{ij}(p_4,p_5) 
 - \sum \limits_{i=1}^{3}  S_{ii}(p_4,p_5) 
\Bigg ] |{\cal M}_{g_1 g_2 \to H g_3}|^2.
\end{split}
\ee
We note that $S_{ij}(p_k)$ is  given in Eq.~(\ref{eq_ijk})
and $S_{ij}(p_4,p_5) $ can be found in Ref.~\cite{lcat3}. 
Using the parametrization of the NNLO phase-space 
and the explicit dependence of the momenta on the singular variable 
$x_1$ that controls the double-soft 
limit, it is straightforward to show that all the singularities 
can be resolved  in triple-collinear sectors. 
In turn, this implies that in such sectors we can 
compute any limit of the form $F(0,x_2,x_3,x_4,...)$ 
from the double soft-limit, with no need
to further distinguish 
the $x_{i=2,.,4}=0$ case from the $x_{i=2,..,4} \ne 0$ one.

Another new element at NNLO is  the  triple-collinear limit.  
Similar to double-collinear limits, 
the  triple-collinear limits are described by the corresponding splitting 
functions. For example, in the case of the  
final-state triple-collinear splitting when the momenta of all final-state 
gluons become parallel, we find 
\be
|{\cal M}_{g_1 g_2 \to H g_3 g_4 g_5}|^2 
\approx 
\frac{g_s^4}{s_{345}^2}
 P^{(g_3g_4g_5)}_{\mu \nu} {\cal M}_{g_1 g_2 \to H g_{345}}^{\mu}  
{\cal M}_{g_1 g_2 \to H g_{345}}^{*,\nu},  
\ee
where $s_{345} = (p_3+p_4+p_5)^2$, 
and $g_{345}$ denotes a gluon with momentum $p_{345} = p_3 + p_4 + p_5$.
The  splitting function $P^{(g_3g_4g_5)}_{\mu \nu}$
was computed in  Ref.~\cite{Catani:1999ss}; it reads
\be\label{pggg}
\begin{split}
& P_{(g_1g_2g_3)}^{\mu \nu} = C_A^2 \Bigg  [
\frac{(1-\ep)}{4 s_{12}^2} \Bigg  [ - g^{\mu \nu} t_{12,3}^2
+ 16 s_{123} \frac{z_1^2 z_2^2}{z_3 ( 1- z_3)} 
\left ( \frac{{\tilde k}_2}{z_2} - \frac{\tilde k_1}{z_1} \right )^\mu
\left ( \frac{{\tilde k}_2}{z_2} - \frac{\tilde k_1}{z_1} \right )^\nu
\Bigg  ]
\\
& - \frac{3}{4}(1-\ep) g^{\mu \nu} + \frac{s_{123}}{s_{12}} g^{\mu \nu} 
\frac{1}{z_3} \left [  \frac{2(1-z_3) + 4 z_3^2}{1-z_3} - \frac{1-2z_3(1-z_3)}{z_1 (1-z_1)} \right ]
\\
&
+ \frac{s_{123}(1-\ep)}{s_{12} s_{13}} 
\Bigg [  2 z_1 \left ( 
{\tilde k_2}^{\mu} {\tilde k_2}^{\nu} \frac{1-2z_3}{z_3(1-z_3)}
+{\tilde k_3}^{\mu} {\tilde k_3}^{\nu}  \frac{1-2z_2}{z_2(1-z_2)} 
\right ) 
\\
& +\frac{s_{123}}{2(1-\ep)}g^{\mu \nu} \left ( 
\frac{4z_2 z_3 + 2z_1(1-z_1) - 1}{(1-z_2)(1-z_3)}
- \frac{1-2z_1(1-z_1)}{z_2 z_3} \right )
\\
& + \left ( {\tilde k}_2^{\mu} {\tilde k}_3^{\nu} + {\tilde k}_3^{\mu} {\tilde k_2}^{\nu} \right ) 
\left ( \frac{2z_2(1-z_2)}{z_3(1-z_3)} - 3 \right ) \Bigg ] \Bigg ] + 5~{\rm permutations}.
\end{split}
\ee
In Eq.~(\ref{pggg}), $s_{ij} = (p_i+ p_j)^2$, $s_{ijk} = (p_i + p_j + p_k)^2$,
and  
$$
t_{ij,k} = 2 \frac{z_i s_{jk} - z_j s_{ik}}{z_i+z_j} 
+ \frac{z_i - z_j}{z_i+z_j}s_{ij}. 
$$
The relevant vectors in this formula are computed in the following way. For final-state triple-collinear splitting, the momentum of the 
resolved gluon $g_3$ defines the collinear direction. The energy fractions 
$z_i$ are obtained  
from energy ratios $z_i = E_{g_i}/E_s$ where  
$E_s = E_{g_3}+E_{g_4}+E_{g_5}$.
Similar to the NLO case, for each phase-space point, 
we compute directions in the plane transverse to 
the collinear direction $p_3$ 
along which collinear limits for $g_4,g_5$ are  taken.  
We denote such directions as $\kappa_{4,5}$, respectively. The vectors 
that enter the triple-collinear splitting function read \footnote{We give 
these vectors for physical labels of the three final state gluons.}
\be
\begin{split} 
& E_{s}^{-1} {\tilde k}_{g_4}^{\mu} = 
z_4 (1-z_4)\kappa_4^{\mu} - z_4 z_5 \kappa_5^{\mu},
\\
& E_{s}^{-1}{\tilde k}_{g_5}^{\mu} 
= z_5 (1-z_5)\kappa_5^{\mu} - z_5 z_4  \kappa_4^{\mu},
\\
& E_{s}^{-1} {\tilde k}_{g_3}^{\mu} = -z_3 z_4 \kappa_4^{\mu} -
  z_3 z_5 \kappa_5^{\mu}.
\end{split} 
\ee
 It is easy to check  that $\sum \limits_{i=3}^{5} {\tilde k}_{g_i} =0$, 
thanks to the energy-conservation condition $\sum \limits_{i=3}^{5} z_i = 1$. 
Finally, we note that for the initial triple-collinear limits 
that correspond to gluons 
$g_4$ and $g_5$ being collinear 
to {\it incoming} gluons $g_1$ or $g_2$,  the above 
formulas are valid up to a replacement $E_{g_3} \to -E_{g_{1}}$ or  $E_{g_3} \to -E_{g_{2}}$.

We
also require the 
 triple-collinear splitting function 
in the strongly ordered configuration,
 $s_{ij}\ll s_{ijk}\ll 1$. 
In principle,  we can obtain it by directly taking the  
limit of Eq.~(\ref{pggg}).
However, it is also easy to compute it directly. 
Indeed, in this case the full triple-collinear $P^{(g_1g_2g_3)}_{\mu\nu}$ 
splitting function factorizes into a 
(spin-correlated) product of ordinary splitting functions. 
We find that in the strongly-ordered $s_{35}\ll s_{345}\ll 1$ limit, the full matrix
element can be written as
\be
|{\cal M}_{g_1 g_2 \to H g_3 g_4 g_5}|^2 
\approx 
\frac{g_s^4}{s_{35} s_{345}} P^{(g_3 g_4 g_5)}_{s.o,\mu\nu}{\cal M}_{g_1 g_2 \to H g_{345}}^{\mu}  
{\cal M}_{g_1 g_2 \to H g_{345}}^{*,\nu}.
\ee
The strong-ordered splitting function reads
\bea
P_{s.o.,(g_3,g_4,g_5)}^{\mu\nu} &=& 16 C_A^2 \left\{
-g^{\mu\nu} \left[ \lp\frac{z_4}{1-z_4} + \frac{1-z_4}{z_4}\rp \lp \frac{z_5}{1-z_5} + \frac{1-z_5}{z_5}\rp + \right.\right. \nn\\
 &+& \left.\left. \frac{z_4}{1-z_4}(1-\ep) 2 z_5(1-z_5) (\kappa_4\cdot \kappa_5)^2\right] 
+ \kappa_4^\mu \kappa_4^\nu (1-\ep) 2 z_4(1-z_4) \times \right. \nn\\
&\times& \lp \frac{z_5}{1-z_5} + \frac{1-z_5}{z_5} + z_5(1-z_5)\rp 
+ \left. \kappa_5^\mu \kappa_5^\nu (1-\ep)2 z_5(1-z_5) \frac{1-z_4}{z_4}
\right\},\nn\\
\eea
where $z_5 = E_{g_5}/(E_{g_3}+E_{g_5})$, $z_4$ is defined as before $z_4 = E_{g_4}/E_s$ and $\kappa_{4,5}$ are the spin-correlation
vectors $\kappa_i = z_i p_3 + k_\perp \kappa_i + y_i \tilde p_3$.

\subsection{Real-virtual corrections} 
\label{lim_rv}

In this Section we consider the  computation of one-loop corrections 
to the real-emission process $g_1g_2\to Hg_3g_4$.  We will refer to this contribution 
as ``real-virtual''.  To calculate this contribution,  we must
integrate the interference of the one-loop
and the tree-level matrix elements  for $g_1g_2 \to H + g_3 g_4$ 
over the NLO phase-space.  The NLO phase-space 
was discussed in Section~\ref{phasespace_nlo}, where we showed how to partition it in such a way 
that soft and collinear singularities  can be extracted.  
Following Section~\ref{phasespace_nlo}, 
we denote the resolved final-state 
gluon as $g_3$ and the potentially 
unresolved final state gluon as $g_4$.  
For each sector, we denote the product of the phase-space parameters 
and the interference of tree- and one-loop matrix elements as 
\be
{\tilde F}_{\rm RV}(x_1,x_2,...)  =  x_1^2 x_2\; 2  
{\rm Re} \left ( {\cal M}_{g_1 g_2 \to H g_3 g_4}^{(1)} 
{\cal M}_{g_1 g_2 \to H g_3 g_4}^{(0),*} \right ),
\label{eq_rv}
\ee
where $x_1$ parametrizes  
the energy of $g_4$ 
and  $x_2 = (1-\cos \theta)/2$ parametrizes the cosine of the 
angle between the direction of the  gluon  $g_4$ and 
the  collinear direction.
This direction is sector-dependent, and is given explicitly later. The ellipses in Eq.~(\ref{eq_rv}) 
stand for other  parameters that are needed 
to fully describe the final-state kinematics.

We must integrate the function ${\tilde F}_{\rm RV}$ 
over the phase-space of the softer gluon; schematically, the integral takes the form 
\be
\int \limits_{0}^{1} 
\frac{{\rm d} x_1 {\rm d} x_2 }{x_1^{1+2\ep} 
x_2^{1+\ep}}{\tilde F}_{\rm RV}(x_1,x_2,..). 
\ee
We note that the extraction of singular limits 
would have been no different from the NLO case discussed in 
Section~\ref{lim_nlo},
if not for the fact that the 
function ${\tilde F}_{\rm RV}$ is not well-defined for $x_1 = 0$ and 
$x_2 = 0$. This happens because ${\tilde F}_{\rm RV}(x_1,x_2,..)$
contains branch cuts in the limits 
$x_1 \to 0$ and $x_2 \to 0$.   To make us of the expansion 
in plus-distributions,  we must isolate 
these  branch cuts before extracting the 
singularities.  We can accomplish this by writing 
${\tilde F}_{\rm RV}(x_1,x_2)$ as  the sum of three terms 
\be
{\tilde F}_{\rm RV}(x_1,x_2,\dots) = 
F_1(x_1,x_2,\dots) 
+ \left ( x_1^2  x_2 \right )^{-\ep} F_2(x_1,x_2,\dots) 
+ x_1^{-2\ep} F_3(x_1,x_2,\dots),
\label{eq_start0}
\ee
where the 
functions $F_{i}(x_1,x_2,\dots)$ 
are free from branch-cut singularities so that their 
values at $x_1=0$ or $x_2=0$ can be computed.  To justify the decomposition of
Eq.~(\ref{eq_start0}), we consider  the limit when the energy of the gluon 
$g_4$ becomes small. 
In this limit, the matrix element squared 
for $g_1g_2 \to H + g_3g_4$ factorizes as~\cite{Catani:2000pi}
\be
|{\cal M}_{g_1g_2\to H g_3g_4} |^2
\approx g_s^2 \mu^{2\ep} 2 C_A 
\left ( 
  {\cal I}_{12,4} 
+ {\cal I}_{13,4} 
+ {\cal I}_{23,4}
 \right ) 
 |{\cal M}_{g_1g_2\to H g_3} |^2, 
\label{eq_start}
\ee
where the soft factors ${\cal I}_{ij,4}$ read 
\be
{\cal I}_{ij,4} = {\cal I}_{ij,4}^{(0)} 
+ 
2 g_s^2 \mu^{2\ep} C_A c_\Gamma {\cal I}_{ij,4}^{(1)}+....
\ee
The function 
${\cal I}_{ij,4}^{(0)}$ is given in Eq.~(\ref{eq_ijk}).
The one-loop function   ${\cal I}_{ij,4}^{(1)}$  reads
\be
{\cal I}_{ij,4}^{(1)} = -\frac{1}{\ep^2} 
\frac{\Gamma^2(1-\ep) \Gamma^2(1+\ep)}{\Gamma(1-2\ep) \Gamma(1+2\ep)}
\left [ S_{ij}(p_4) \right ]^{\ep} S_{ij}(p_4),
\ee
where  the eikonal factor  $S_{ij}$ can be found in  Eq.~(\ref{eq_ijk}).

We next expand  Eq.~(\ref{eq_start}) through first order in the strong coupling constant to obtain 
\be
\label{soft_limit_rv}
\begin{split} 
 &  2  {\rm Re} \left (   {\cal M}^{(1)}(g_1,..,g_4) {\cal M}^{(0),*}(g_1,..,g_4) \right ) 
\Bigg |_{g_4 \to 0}
 \to 
\\
 & 4 g_s^2 c_\Gamma C_A   
\Bigg (
\left [ 
{\cal I}_{12,4}^{(0)} 
+ {\cal I}_{13,4}^{(0)} 
+ {\cal I}_{23,4}^{(0)}  \right ]
2  {\rm Re} \left ( {\cal M}_{g_1g_2 
 \to Hg_3}^{(1)} {\cal M}_{g_1g_2 \to Hg_3}^{(0),*}(g_1,..,g_3) \right )   \\
& + \left [  {\cal I}_{12,4}^{(1)} + {\cal I}_{13,4}^{(1)} 
+ {\cal I}_{23,4}^{(1)}  \right ]|{\cal  M}_{g_1g_2 \to Hg_3}^{(0)}|^2
\Bigg ).
\end{split}
\ee
Using the
explicit expression for the function $S_{ij}$, 
and the parametrization of $p_4$ in terms of $x_1$ 
and $x_2$ for a given collinear direction, 
it is easy to verify that 
when the collinear direction is  the 
direction of the hard gluon $g_h$,  
$h \in (1,2,3)$,  terms in Eq.~(\ref{soft_limit_rv}) that are 
proportional to ${\cal I}^{(0)}_{ij,4}$ 
contribute to $F_1$, terms proportional to 
${\cal I}_{ih,4}^{(1)}$ or 
${\cal I}_{hi,4}^{(1)}$  contribute to 
$F_2$ and terms that are proportional to ${\cal I}_{ij,4}^{(1)}$ with 
$i \ne h, j\ne h$, contribute to $F_3$. 

As the next step, we check that the parametrization in 
Eq.~(\ref{eq_start0}) 
is consistent with 
the behavior of the real-virtual 
matrix elements in the collinear limit.  Consider for 
definiteness the 
${\rm Sc}^{(43)}$ sector, where the singularity occurs 
when gluon $g_4$ becomes collinear to gluon $g_3$. 
In the collinear limit, color-ordered matrix elements 
factorize as follows~\cite{Kosower:1999rx}
\be
{\cal M}_{g_1g_2 \to Hg_3g_4}^{(1)} = g_s {\rm Split}_{{\tilde g}_3 \to g_3g_4}^{\rm tree} 
\otimes {\cal M}_{g_1g_2 \to H {\tilde g}_3}^{(1)} 
+ g_s^3 {\rm Split}_{{\tilde g}_3 \to g_3g_4}^{1\rm loop} \otimes M_{g_1g_2 \to H {\tilde g}_3}^{(0)},
\ee
where the convolution sign refers to a sum over the helicities of the intermediate gluon 
${\tilde g}_3$.  The 
tree splitting function for $g_p \to g_ag_b $  reads~\cite{Kosower:1999rx} 
\be
{\rm Split}^{\rm tree}( g_p \to g_ag_b,z) = 
-\frac{\sqrt{2}}{{s_{ab}}} \left ( -\epsilon_a \cdot \epsilon_b k_b \cdot \epsilon_{p} 
+ k_b \cdot \epsilon_a \epsilon_p \cdot \epsilon_b 
- k_a \cdot \epsilon_b \epsilon_a \cdot \epsilon_p 
\right ),
\ee
where by definition $z=E_a/(E_a+E_b)$ is the momentum fraction carried by $g_a$,  
$\epsilon_{a,b,p}$ are the polarization vector of  three particles 
that participate in the splitting and all momenta are taken to be {\it outgoing}, so that $p+k_a + k_b = 0$. 
The one-loop splitting function is given by 
\be
\begin{split} 
& {\rm Split}^{1-\rm loop }( g_p \to g_a g_b,z) =
\frac{1}{2}~F_{\rm tree}(z)~{\rm Split}^{\rm tree}(g_p \to g_a g_ b,z) \\
& + \frac{1}{\sqrt{2} s_{ab}^2}F_{\rm new}(z) \left ( k_a - k_b \right ) \cdot \epsilon_1 
\left ( s_{ab} \epsilon_a \cdot \epsilon_b - 2 k_b \cdot \epsilon_a k_a \cdot \epsilon_b \right ).
\end{split} 
\ee
In conventional  dimensional  regularization, 
the functions $F_{\rm tree}$ and $F_{\rm new}$ read 
\be
\begin{split} 
& F_{\rm tree} = \frac{1}{2} \left ( \frac{\mu^2}{-s_{ab}} \right )^{\ep} 
\left [ z f_1(z) + (1-z) f_1(1-z) - 2 f_2 \right ], 
\\
& F_{\rm new} = \frac{\epsilon^2(1-\epsilon)}{(1-2\epsilon)(3-2\epsilon)} 
\left ( \frac{\mu^2}{-s_{ab}} \right )^{\ep} f_2, 
\end{split} 
\label{eqs_f}
\ee
where 
\be
\begin{split} 
& f_1 = \frac{2}{\epsilon^2} c_\Gamma \left ( - \Gamma(1-\ep) \Gamma(1+\ep) z^{-1-\ep} (1-z)^{\ep} 
- \frac{1}{z} + \frac{(1-z)^\ep}{z} ~_2 F_{1}(\ep,\ep,1+\ep,z) \right ), \\
&f_2 = -\frac{1}{\epsilon^2} c_\Gamma.
\end{split} 
\label{eqs_f1}
\ee
In calculating the splitting functions, one has to be careful with imaginary parts. 
Note that $s_{ab}$ in Eq.(\ref{eqs_f}) can be both positive 
and negative so that $(-s_{ab})^{-\ep}$ 
may or may not have an imaginary part. Also, because of analytic continuation in the 
space-like region $z$ can be smaller or larger 
than one,  in which case $f_1$ may have an imaginary part.  However, these cases are mutually exclusive since $s_{ab}$ is positive 
for the final state splitting where $z < 1$, 
and negative for the initial state splitting where $z>1$. 
As a result, we do not need to care about the interference 
of two imaginary parts, or about their consistent definition.
We require the 
expansion of the hypergeometric function in Eq.(\ref{eqs_f1}) 
through ${\cal O}(\ep^{3})$.  It reads 
\be
\begin{split} 
~_2F_{1}(\ep,\ep,1+\ep,z) & = 
1+{\rm Li}_2(z)\ep^2
+ \ep^3 \Bigg [  
\zeta_3 + \frac{1}{2}\ln z  \ln^2(1-z)+\ln(1-z) {\rm Li}_2(1-z)
\\
& 
 -{\rm Li}_3(1-z)-{\rm Li}_3(z) \Bigg  ] + {\cal O}(\ep^4). 
\end{split} 
\ee

We will need products of splitting amplitudes summed over polarization 
states of unresolved particles. These polarization states must be taken 
in $d$-dimensions but, because of the real-virtual kinematics, 
the four-momenta of all gluons are four-dimensional.   
We write these products as 
\be
 \sum_{\lambda_a,\lambda_b} 
{\rm Split}^{\rm tree}(g_p^{\mu} \to g_a g_b) {\rm Split}^{\rm tree}(g_p^{\nu} \to g_a g_b)
= \frac{2}{s_{ab}} P^{(gg),\mu \nu}(z,\kappa_a,\ep),
\ee
with the LO splitting function defined in Eq.~(\ref{Pgg0munu})
and 
\be
\begin{split} 
& \sum_{\lambda_a,\lambda_b} 
{\rm Split}^{\rm tree}(g_p^{\mu} \to g_a g_b) 
{\rm Split}^{1-\rm loop }(g_p^{\nu} \to g_a g_b)
=  \frac{1}{s_{ab}} P_{gg,\rm int}^{\mu \nu}(z,\kappa_a,\ep),\\ 
& 
P_{gg,\rm int}^{\mu \nu}(z,\kappa_a,\ep) = F_{\rm tree}(z) P^{(gg),\mu \nu} (z,\kappa_a,\ep)
- 2 F_{\rm new}(z) ( 1- 2z(1-z)\ep) \kappa_a^\mu \kappa_a^\nu.
\end{split} 
\ee

With these definitions, we are in position to present 
the limiting behavior of the interference of one-loop and tree amplitudes 
in the collinear limit. We find 
\be
\begin{split} 
& 2 {\rm Re} \left ( {\cal M}^{(0),*}_{gg \to Hgg} {\cal M}_{gg \to Hgg}^{(1)} \right ) 
= \frac{2}{s_{ab}}{\rm Re} \left ( {\cal M}^{(0),*}_{\mu,gg \to Hg} {\cal M}_{\nu, gg \to Hg}^{(1)} \right ) 
2 P^{(gg),\mu \nu}
\\
& +
\frac{2}{s_{ab}} {\rm Re} \left ( {\cal M}^{(0),*}_{\mu,gg \to Hg} {\cal M}_{\nu,gg \to Hg}^{(0)} \right ) 
 {\rm Re} \left ( P_{gg,\rm int}^{\mu \nu} \right ). 
\end{split} 
\label{eq416}
\ee
Note that in the last term we have taken the interference splitting function 
outside of the real part and have replaced this splitting function by its real part.  
We are allowed to do that because
$P_{gg,\rm int}^{\mu \nu}$ is a symmetric tensor, so that we can write  
\be
\begin{split} 
{\cal M}^{(0),*}_{\mu,gg \to Hg} {\cal M}_{\nu,gg \to Hg}^{(0)}  P_{gg,\rm int}^{\mu \nu}
& =  \frac{1}{2}\left ( 
{\cal M}^{(0),*}_{\mu,gg \to Hg} {\cal M}_{\nu,gg \to Hg}^{(0)} + (\mu \leftrightarrow \nu)
\right ) P_{gg,\rm int}^{\mu \nu} 
\\
& = {\rm Re} \left ( {\cal M}^{(0),*}_{\mu,gg \to Hg} {\cal M}_{\nu,gg \to Hg}^{(0)} \right ) 
P_{gg,\rm int}^{\mu \nu}.
\end{split}
\ee
This observation  is useful since 
we need $P_{gg,\rm int}^{\mu \nu}$ for $z$ both smaller and larger than one, 
and   the above equation implies that the analytic continuation 
of $P_{gg,\rm int}^{\mu \nu}(z)$ can be done in an 
arbitrary way since the imaginary part drops out.

It follows from Eq.~(\ref{eq416})  
that in the collinear limit the amplitude 
has a single branch cut   $x_2^{-\ep}$. 
Indeed,  $P_{gg}^{\mu \nu}(z)$ is a rational polynomial 
of $z$ and therefore contributes 
to $F_1$ while $P_{gg,\rm int}^{\mu \nu}$ is proportional 
to $s_{ab}^{-\ep} \sim x_2^{-\ep}$.  We therefore 
match the $x_2 \to 0$ limit 
to  $F_1$ and $F_2$ and require that $F_3(x_1,0)$ 
vanishes.
Finally, we note that 
the splitting functions in 
Eq.~(\ref{eq416}) exhibit spin correlations; we can handle them in 
exactly the same way as described 
in the Section dedicated to next-to-leading order computations, where 
we explained that for each 
phase-space point we compute the vector $\kappa^\mu$ such that 
$k_\perp^{\mu} = \sqrt{-k_\perp^2} \kappa^\mu$. We do this at the level 
of phase-space point generation, where  
we resolve all singularities related to the collinear 
$k_\perp^2 \to 0$ limit analytically.  Once the
vectors $\kappa_\mu$ are known, we can 
rewrite Eq.~(\ref{eq416}) through sums over ($d$-dimensional) 
helicities in complete analogy with what was done at next-to-leading order.

\section{Higher-order $\epsilon$ terms in amplitudes}
\label{ampl_ep}

The computational algorithm that we discuss in this paper 
is based on conventional dimensional regularization, in which the polarization 
states of all particles are continued to $d$-dimensions. Therefore, 
it becomes  an important issue  in our construction to understand 
how scattering-amplitude contributions at higher orders in $\epsilon$ can be calculated. The goal of this 
Section is to discuss this issue.

We  begin by pointing out that the highest-multiplicity 
amplitudes at any order of perturbation theory are needed 
to ${\cal O}(\ep^0)$, since they only contribute to the finite parts 
of the relevant correction. 
Therefore, at NNLO for Higgs plus jet production, 
we can use four-dimensional expressions for  tree amplitudes 
$gg \to H g g g$ and  we can truncate one-loop 
amplitudes $g g \to H g g$ at ${\cal O}(\epsilon^0)$. 
However, for lower-multiplicity amplitudes, 
such as tree-level $gg\to Hg$ and $gg\to Hgg$, 
we need to  know higher-order $\ep$ terms.
In principle, it can be expected that 
higher-order $\ep$ terms for the one-loop $gg\to Hg$ 
amplitudes are needed but, as it was pointed out in 
Ref.~\cite{Weinzierl:2011uz}, 
this is not the case. Indeed, the 
 ${\cal O}(\ep)$ contributions to these amplitudes cancel out 
between the two-loop virtual 
correction, the square of the one-loop amplitude and the singular limit of 
the real-virtual correction.   We use this cancellation 
as a  consistency check on our numerical implementation.  
We calculate  the one-loop $gg \to Hg$ amplitude 
through ${\cal O}(\ep^2)$ and check that the
higher-order $\ep$ terms 
do not contribute to  the final result due to the
above-mentioned cancellations. 

We note that computations of matrix elements squared 
for $d \ne 4 $  are  straightforward if they  are  performed by adding 
and squaring Feynman diagrams.  All one needs to do in this 
case is to use the correct contractions of metric tensors  obtained 
after summing over polarization states of external particles. 
Unfortunately, if a calculation is done in this way, the results 
rapidly become unwieldy, especially when a large number of 
gluons is involved.  Instead,  we decided to compute higher-order $\ep$ terms 
directly  at the {\it amplitude level}. This is possible because 
for lower-multiplicity  final states  we can choose to parametrize 
momenta of all particles as four-dimensional. 
The 
``extra-dimensional'' polarization 
vectors then have a simple property that they are orthogonal 
to all momenta,  $\epsilon_s \cdot p_i = 0$. Therefore, the only 
way such polarizations can contribute to the amplitude and give 
non-vanishing contributions is through scalar products $\epsilon_{i,s} \cdot 
\epsilon_{j,s'} = -\delta_{s,s'}$. 
This implies that  the necessary condition for 
the amplitude to be non-vanishing is that an
{\it even} number of particles has 
``extra-dimensional'' polarizations. 

To illustrate how this works in detail, 
we consider the tree-level $gg\to Hg$ amplitudes. We write the full
amplitude as
\be
\mathcal A(1^{h_1},2^{h_2},3^{h_3}) = 2 i \lambda^{(0)}_{Hgg} g_s (F^{c_2})_{c_1 c_3} A(1^{h_1},2^{h_2},3^{h_3}),
\ee
where $\lambda^{(0)}_{Hgg}$ is the Higgs effective coupling defined in Eq.~(\ref{lambdadef}) and $(F^{c_2})_{c_1c_3}=
-i\sqrt{2} f^{c_1c_2c_3}$ is the color generator in the adjoint representation.
Apart from normal helicities, we can
have amplitudes where exactly one  pair of gluons has {\it identical}   
extra-dimensional polarizations. 
We find  one such independent color-ordered amplitude which reads 
\be
A(1^s,2^s,3^+) = 
\frac{[1 3] [3 2]}{[2 1]}+m_h^2 \frac{\langle 1 2\rangle}{\langle 2 3\rangle  \langle 3 1\rangle }.
\ee
Because extra-dimensional polarizations in the above amplitude 
should be the same 
and because there are $d-4=-2\ep$ extra-dimensional directions, 
the matrix element squared for $gg \to Hg$ can be written as
\be
\begin{split} 
\left| M_{gg \to Hg} \right|^2 = & \sum_{h_i=\pm} \left | 
\mathcal A(1^{h_1},2^{h_2},3^{h_3})\right|^2 
\\
& -2\ep \lp \sum_{h=\pm} \left| \mathcal A(1^{h},2^s,3^s)\right|^2 
+ \left| \mathcal A(1^s,2^h,3^s)\right|^2 
+ \left| \mathcal A(1^s,2^s,3^h)\right|^2 \rp
\\
= & 
(1-\ep) \sum_{h_i=\pm} \left | 
\mathcal A(1^{h_1},2^{h_2},3^{h_3})\right|^2
-4 \ep m_h^2.
\end{split}
\ee
The final result here can be easily verified since it implies that, 
except for the $-4 m_h^2$ term, the ${\cal O}(\ep)$ contribution 
to the squared matrix element for $gg \to Hg$  and the
${\cal O}(\ep^0)$ contribution coincide up to a sign. 

We note that, in addition to the matrix element squared, 
our construction requires more complicated objects that 
appear in collinear limits
\be
|{\cal M}_{\rm spin}(n)|^2 = 
\sum_{h_2,h_3}  A(1^{n},2^{h_2},3^{h_3}) \; 
A^*(1^{n},2^{h_2},3^{h_3}), 
\ee
where the amplitudes on the right-hand side are computed under the 
assumption that the polarization vector of the gluon $g_1$ is $n^\mu$.
To calculate $|{\cal M}_{\rm spin}(n)|^2$, we write $n$ as a linear 
combination of suitable polarization vectors of gluon $g_1$. In doing 
so, it is important to remember that the vector $n$ can have extra-dimensional 
components. We write 
\be
\begin{split} 
& |{\cal M}_{\rm spin}(n)|^2 = \sum \limits_{h_1,h_1'}^{}
\rho(n,h_1,h_1') |{\cal M}_{\rm spin}(h_1,h_1')|^2,
\\
& |{\cal M}_{\rm spin}(h_1,h_1')|^2 = 
\sum \limits_{h_2,h_3}^{} 
    A(1^{h_1},2^{h_2},3^{h_3}) \; 
 A^*(1^{h_1'},2^{h_2},3^{h_3}), 
\end{split} 
\ee
where $\rho(n,h,h') = (n \cdot \epsilon_h) \; (n \cdot \epsilon_{h'}^*)$.
The helicity labels $h_1,h_1'$ can assume the following 
values:  $(h_1,h_1') = (ij), (i,s), (s,j), (s,s)$
where $i,j = \pm$.  It is straightforward to compute 
$|{\cal M}_{\rm spin}(h_1,h_1')|^2$ for all pairs of helicity 
labels. The key point here is that, for non-vanishing 
amplitudes, there must be either zero or two $s$-helicity labels. 
We are then left with the following non-zero entries 
\be
\begin{split} 
& \left | M_{\rm spin}(i,j) \right|^2
=   \left|M_{\rm spin} \right|^2_{d=4}(i,j)
-2\ep \left[\mathcal A(1^i,2^s,3^s) \mathcal A^*(1^j,2^s,3^s)\right],\\
&  \left|M_{\rm spin} (s,s) \right|^2 = 
\sum_i \left[\mathcal A(1^s,2^i,3^s) \mathcal A^*(1^s,2^i,3^s) + 
\mathcal A(1^s,2^s,3^i) \mathcal A^*(1^s,2^s,3^i)\right].
\end{split} 
\ee
Finally, we note that following a similar approach, 
it is straightforward  to obtain 
the double-correlated matrix element $|M_{\rm spin}(h_1,h_1',h_2,h_2')|^2$, 
which is needed to describe singular limits in the double-collinear sectors.

We also need to discuss the $\ep$-dependent parts of  
$0 \to Hgggg$ amplitudes. In this case, we use the following color 
decomposition
\be
\mathcal A(1^{h_1},2^{h_2},3^{h_3},4^{h_4}) =2 i \lambda^{(0)}_{Hgg} g_s^2\sum_{\sigma\in S_2}
(F^{c_{\sigma(2)}} \cdot F^{c_{\sigma(3)}})_{c_1 c_4}  A(1^{h_1},2^{h_2},3^{h_3},4^{h_4}).
\ee
The situation now 
is slightly more involved than before because there are more options 
for extra-dimensional polarizations. Indeed, with four gluons 
the amplitude does not vanish if all of them have identical extra-dimensional
polarizations but also  when there are two pairs of gluons with different 
extra-dimensional polarizations.  
We will  denote color-ordered amplitudes for these cases as 
$A(1^s,2^s,3^s,4^s)$ and $ A(1^s,2^s,3^{s'},4^{s'})$.   These amplitudes can be written in a relatively compact form. For 
example, 
\be
\begin{split} 
& A(1^s,2^s,3^s,4^s) = 
\sum \limits_{i=0}^{4} {\cal R}^{i} F(1,2,3,4),
\\
& F(1,2,3,4) =
\frac {m_h^2} {s_{123}} 
  \lp 1 + \frac{s_{12}}{s_{23}}+\frac{s_{23}}{s_{12}}\rp 
- \lp \frac{m_h^2}{2s_{12}}+\frac{m_h^2}{2s_{23}}\rp 
+  \frac{1}{2}\lp \frac{s_{12}s_{34}}{s_{14}s_{23}} 
- \frac{s_{13}s_{24}}{s_{12}s_{34}} \rp, 
\\
& A(1^s,2^s,3^{s'},4^{s'}) = m_h^2
  \left( \frac{s_{14} +s_{12}}{s_{12}s_{124}}
    -\frac{s_{13}}{s_{12}
   s_{123}}+\frac{s_{14}+s_{34}}{s_{34}s_{134}}
           -\frac{s_{24}}{s_{34}
   s_{234}}  \right)
\\
& \;\;\;\;\;\;\;\;\;\;\;\;\;\;\;\;\;
+\frac{s_{14} s_{23}}{s_{12}
   s_{34}}-\frac{s_{13} s_{24}}{s_{12} s_{34}}-1,\\
& A(1^s,2^{s'},3^s,4^{s'}) =  2- m_h^2
   \left( \frac{1}{s_{124}}+\frac{1}{s_{134}}+\frac{1}{s_{234}}+
   \frac{1}{s_{123}}\right),
\end{split} 
\ee
where ${\cal R}$ is a permutation  operator defined as 
${\cal R}F(a,b,c,d) = F(b,c,d,a)$.  The amplitudes remain compact even if only one 
 pair of gluons has extra-dimensional 
polarization.  For example, we obtain
\be
A(1^s,2^+,3^s,4^+) = -
\frac{\langle 1 |p_h |4] \langle
   3|p_h|4]}{s_{123} \langle 12\rangle  \langle
   23\rangle }+\frac{\langle 1|p_h|2] \langle
   3|p_h|2]}{s_{134} \langle 14\rangle  \langle
   34\rangle }+\frac{m_h^2 \langle 13\rangle ^2}{\langle
   12\rangle  \langle 14\rangle  \langle 23\rangle  \langle
   34\rangle },
\ee
where $p_h$ is the outgoing momentum of the Higgs boson.  Similar results for all other helicity configurations
can be derived.

We are now in position 
to discuss how to use these amplitudes to assemble 
the matrix element squared for $0 \to Hgggg$, 
summed over polarization vectors of 
all gluons. 
Similar to the $0 \to Hggg$ case that we already discussed, 
amplitudes with two gluons with extra-dimensional polarizations, 
e.g. $A(1^i,2^j,3^s,4^s)$,  enter with a $(d-4) = -2\ep$ weight.
The same is true for the amplitude 
$A(1^s,2^s,3^s,4^s)$, as $s$ just counts the number of extra-dimensional 
polarizations. 
For amplitudes like $A(1^s,2^s,3^{s'},4^{s'})$, we have again 
$d-4$ polarizations
for the index $s$ and $d-5$ for $s'$ since,   by construction,  $s\ne s'$. 
Combining everything, we obtain
\be
\label{me2ep_gggg}
\begin{split}
& |M(H,g_1,g_2,g_3,g_4)|^2 = |M(H,g_1,g_2,g_3,g_4) |^2_{d=4} 
- 2 \ep \bigg[ |\mathcal A(1^s,2^s,3^s,4^s)|^2  
\\
& +  \sum_{h_i,h_j}\lp
  |\mathcal A(1^s,2^s,3^{h_i},4^{h_j})|^2 
+ |\mathcal A(1^s,2^{h_i},3^s,4^{h_j})|^2 
+ |\mathcal A(1^s,2^{h_i},3^{h_j},4^s)|^2\right. 
\\
& \left.\left. + |\mathcal A(1^{h_i},2^s,3^s,4^{h_j})|^2 +
|\mathcal A(1^{h_i},2^s,3^{h_j},4^s)|^2 
+ |\mathcal A(1^{h_i},2^{h_j},3^s,4^s)|^2 \rp \right] + \\
& + 2\ep(2\ep+1) \left[ |\mathcal A(1^s,2^s,3^{s'},4^{s'})|^2 + 
|\mathcal A(1^s,2^{s'},3^{s},4^{s'})|^2 +
|\mathcal A(1^s,2^{s'},3^{s'},4^{s})|^2\right].
\end{split} 
\ee

In full analogy with the amplitudes for $0 \to Hggg$, we can calculate 
$|M(h_1,h_1')_{\rm spin} |^2$ which is required to describe spin 
correlations in collinear limits.  A simple analysis reveals that 
this spin-correlated matrix element squared is non-vanishing 
provided that  $(h_1,h_1')=(i,j)$ or $(h_1,h_1')=(s,s)$ so 
that no mixed terms as $(i,s)$ or $(s,s')$ appear. 
The result can be written as 
\be
\begin{split} 
&  |M(i,j)_{\rm spin} |^2
 =  |M(i,j)_{\rm spin} |^2_{d=4}
 - 2\ep \sum_{h =\pm} \left ( 
\mathcal A(1^{i},2^s,3^3,4^h) \mathcal A^*(1^{j},2^s,3^s,4^h) 
\right.
\\
& \left. 
+\mathcal A(1^{i},2^s,3^h,4^s) \mathcal A^*(1^{j},2^s,3^h,4^s)
+\mathcal A(1^{i},2^h,3^s,4^s) \mathcal A^*(1^{j},2^h,3^s,4^s) \right ),
\\
& |M(s,s)_{\rm spin}|^2  = \mathcal A(1^s,2^s,3^s,4^s) 
\mathcal A^*(1^s,2^s,3^s,4^s) 
+ \sum_{i,j}  \left ( 
 \mathcal A(1^s,2^i,3^j,4^s )  \mathcal A^*(1^s,2^i,3^j,4^s)
\right.
\\
& \left. 
+\mathcal A(1^s,2^i,3^s,4^j )  \mathcal A^*(1^s,2^i,3^s,4^j)
+\mathcal A(1^s,2^s,3^i,4^j )  \mathcal A^*(1^s,2^s,3^i,4^j)
\right )
\\
& -(1+2\ep)\left (
\mathcal A(1^s,2^s,3^{s'},4^{s'} )  \mathcal A^*(1^s,2^s,3^{s'},4^{s'})
+\mathcal A(1^s,2^{s'},3^{s},4^{s'} )\mathcal A^*(1^s,2^{s'},3^s,4^{s'})
\right.
\\
&\left. 
+\mathcal A(1^s,2^{s'},3^{s'},4^s )  \mathcal A^*(1^s,2^{s'},3^{s'},4^s)
\right ).
\end{split} 
\ee

\section{Numerical implementation}
\label{sec:num}

In this Section we discuss the implementation of the algorithm 
described  above in a numerical program. We choose to do so 
in FORTRAN 90 since it offers the option  of  performing 
computations in double- and quadruple precision in a 
straightforward way. Such  flexibility is important 
because  in our  framework  singular limits are approached numerically,
and  we have to find a balance between the speed of the code and 
the numerical stability which requires switching to quadruple 
precision computations when close to singularities.  

For numerical implementation of the required amplitudes 
we used, as much as possible, pieces 
of the FORTRAN 77 code MCFM \cite{mcfm}. After translating  
to FORTRAN 90 we checked our numerical implementation 
of the tree-level amplitudes for $0 \to Hggg$, $0 \to Hgggg$ and $0 \to Hggggg$ 
processes  against MadGraph~\cite{Alwall:2011uj}.  As we explained earlier, 
since we work in conventional dimensional 
regularization, we need to know ${\cal O}(\ep)$ parts of 
tree-level amplitudes, which are presented in the previous Section. 
These ${\cal O}(\ep)$ parts were checked against a Feynman diagram-based 
computation of amplitudes squared where explicit sums over gluon polarizations
were performed.   The relevant diagrams  for $0 \to H+ng$, $n=3,4$ 
were  obtained with  QGRAF~\cite{Nogueira:1991ex} 
and manipulated with FORM~\cite{Vermaseren:2000nd}. 
Finally, we note that since we require the one-loop corrections to 
$gg \to Hg$ through ${\cal O}(\ep^2)$, we recomputed the
one-loop $gg\to Hg$ amplitudes and compared them against 
the results presented in~\cite{Schmidt:1997wr}.
For the $0 \to Hggg$ one-loop amplitudes, we 
borrowed significant parts of the FORTRAN code from MCFM.
The one-loop integrals that are required for this calculation are 
computed using QCDloops~\cite{Ellis:2007qk}.
The box one-loop master integral $gg \to Hg$ is needed to higher 
orders in the expansion in $\epsilon$, and can be obtained 
starting from  an all-orders result in Ref.~\cite{Anastasiou:2011qx}.

A central part of the described computational algorithm is the calculation 
of integrals of the following form 
\be
\int \limits_{0}^{1} {\rm d} x_1..{\rm d}x_n \; {\rm d} \vec y \; 
{\cal D}_{i_1}(x_1)...{\cal D}_{i_n}(x_n) F(x_1,..x_n, \vec y),
\label{eq_77}
\ee
where $n$ counts the number of singular phase-space variables (
$n=4$ for double-real and $n=2$ for real-virtual), 
$\vec y$ collectively denotes all non-singular variables, the functions 
${\cal D}(x)$ are defined as 
\be
{\cal D}_0(x) = \delta(x),\;\;\;\;\;
{\cal D}_{i}(x) = \left [ \frac{\ln^{i-1} x}{x} \right ]_+,\;\;\;
\ee
and $\sum_j  i_j \le 3$.
The function $F(x_1,..x_n,\vec y)$ is 
obtained by multiplying the matrix element 
squared for a particular physics process by appropriate powers of $x_1,..x_n$, 
as explained in Section~\ref{phasespace}.  To compute multi-dimensional integrals of the type  
shown in Eq.~(\ref{eq_77}), we use the adaptive Monte-Carlo 
algorithm VEGAS \cite{lepage} as implemented  in the CUBA library 
\cite{hahn}. We note that when plus distributions are expanded out
in Eq.~(\ref{eq_77}), we obtain  integrands that 
are iterations of the following basic form 
\be
x_i^{-1} \left [ F(x_1,\dots x_{i-1},x_i,x_{i+1} \dots ) 
- F(x_1,\dots x_{i-1},0,x_{i+1}, \dots ) \right ].
\label{eq_78}
\ee
To  understand subtleties of the numerical implementation 
of  Eq.~(\ref{eq_77}), it is 
important 
to realize that the two terms in the numerator of Eq.~(\ref{eq_78})  
are computed differently in the numerical code.
Indeed, the function  $F(x_1,\dots x_{i-1},x_i,x_{i+1} \dots )$
is calculated  from the matrix element squared that describes the 
highest-multiplicity process for a given channel (for example, 
it is   $0 \to Hggggg$ for the double-real emission processes). 
On the other hand, 
$F(x_1,\dots x_{i-1},0,x_{i+1} \dots )$ is computed by first 
analytically calculating the appropriate singular limit from 
the full matrix element and then 
implementing that limit as an independent function or subroutine 
in the numerical code.  This implies two things. First, 
\be
\lim_{x_i \to 0} F (..x_i,...) = F(..,0,..),
\label{eq72}
\ee
is an important and non-trivial check  of  the calculation 
and of its implementation in  the numerical program. Second, 
because the full matrix elements become numerically unstable for 
very small values of $x$, it is not possible to calculate 
integrands  as in Eq.~(\ref{eq_77}) all the way to 
$x_{1,..,n}= 0$.  In our numerical implementation, we follow 
the approach of Ref.~\cite{Czakon:2011ve} and 
require that the product of all generated singular variables
is larger than a small parameter $\delta_c$, 
\be
x_1 x_2 .. x_n \ge \delta_c.
\ee
Independence of the final result from the value 
of $\delta_c$  is an essential check of the correctness 
of the numerical implementation; we will discuss the necessary 
condition for that in the next Section.  

To obtain  results for partonic cross-sections that will be presented in the 
next Section, we use $\delta_c = 10^{-10}$. We introduce a switch 
in the program that forces a quadruple precision calculation of the 
integrand in Eq.~(\ref{eq_77}) to occur provided that 
$x_1 x_2 .. x_n \le \delta_s$, where $\delta_s$ is conservatively  chosen to 
be $\delta_s = 10^{-7}$.  We find that, compared to a pure double-precision 
computation, our implementation of the  switch  slows  a 
calculation by about a factor of two. 
This, however, is not  a problem since the program is  quite fast because 
it employs helicity amplitudes 
to construct the relevant matrix elements.  To illustrate 
how fast the program is, we note that 
to get contributions  to integrated 
partonic cross-sections for one center-of-mass collision energy, 
we need about half an hour to obtain all the poles in $\epsilon$ 
and about four hours to obtain all the relevant finite parts provided 
that a calculation is done on a cluster of twenty eight-core 
$2.83~{\rm GHz}$ nodes.

Finally, we note that our current implementation of the numerical integration 
procedure allows us to calculate partonic cross-sections but  not kinematic 
distributions. This, however, seems a relatively minor problem  since 
at every step of the calculation we know the kinematics of the final 
state and we can access the weight. It appears therefore  that the current 
implementation can be easily extended to make a true parton-level 
generator capable of computing different observables in a single run.
In fact, the possibility to do this within the current framework 
was recently demonstrated for the simpler processes 
$t \to be^+\nu$ and $b \to u e \bar \nu$ 
in Refs.~\cite{Brucherseifer:2013iv,Brucherseifer:2013cu}. 
We plan to return to the discussion of this issue in the context of 
Higgs boson production in the near future. 

\section{Checks and final results}
\label{sec:checks}

In this Section, we describe checks on the calculation and 
present the results for the partonic cross-section. 
The first check that we describe follows from the fact that, 
in the numerical program, Eq.~(\ref{eq72}) 
is non-trivial to satisfy, because of the different ways in which the function 
$F(x_1,x_2,..)$ and its boundary values are computed.  On 
the other hand, Eq.~(\ref{eq72})  is a necessary requirement 
for the existence of integrals shown 
in Eq.~(\ref{eq_77}), so that its validity in our numerical program 
should be carefully investigated.  To check Eq.~(\ref{eq72}), we compute 
\be
L_{i_1,i_2,...}(t) = 1 
- \frac{F(x_1,..t x_{i_1},..,t x_{i_2},.. )}
{F(x_1,..,0,x_{i_1+1}..0,x_{i_2+1})},
\ee
as a function of $t\to 0$, for  random choices 
of $\vec x = [x_1,x_2,...]$.  If the calculations are done properly, we should 
find 
$L_{ij..}(t) \to 0$ as $t \to 0$, independent of $\vec x$. Also, 
because the variables $x$ define the kinematics of the process, 
each function $L_{i_1,i_2..}(t)$ probes a particular singular limit of the 
full amplitude. 

\begin{figure}[!h]
\includegraphics[angle=-90,width=0.5\textwidth]{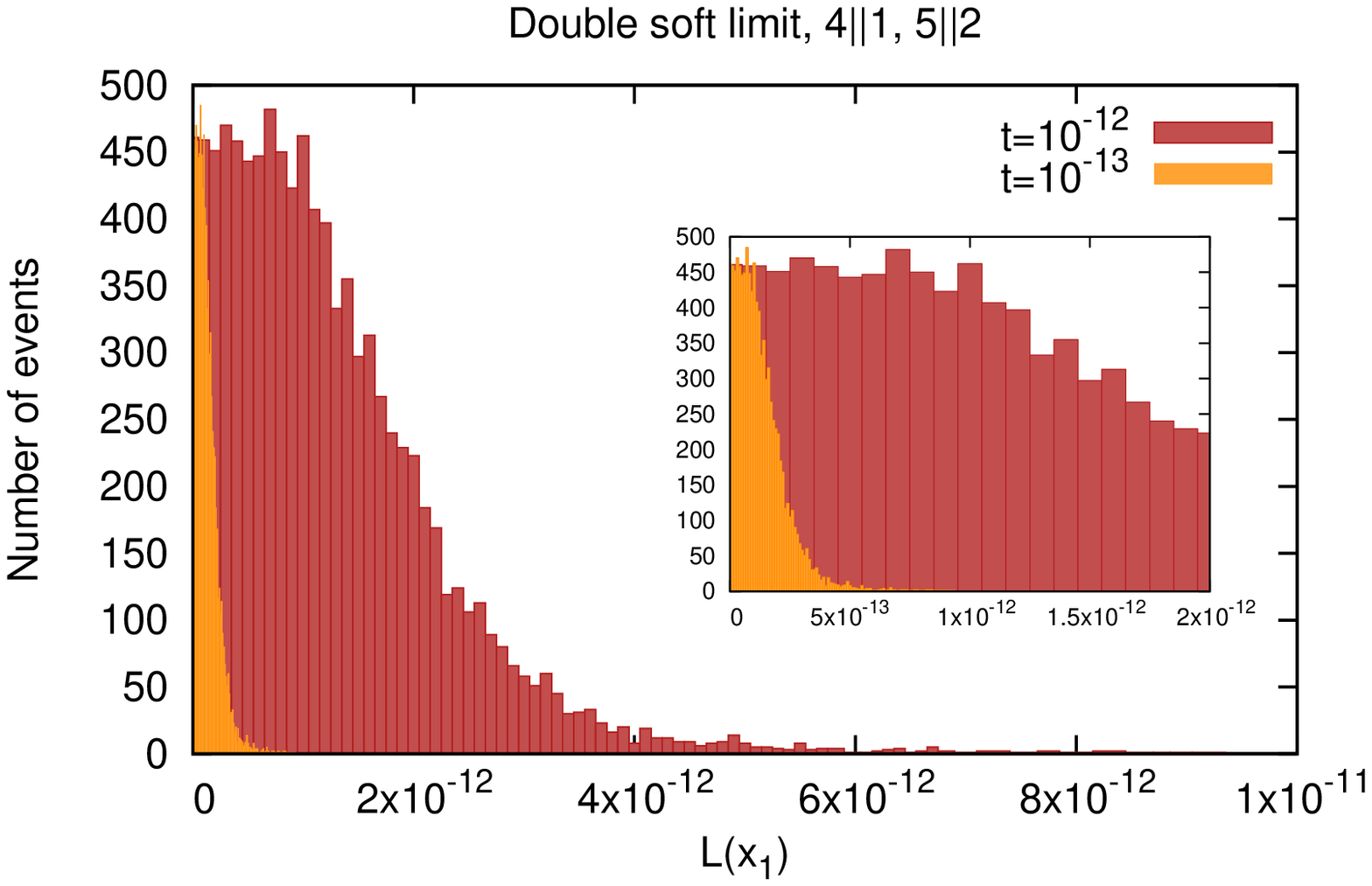}
\includegraphics[angle=-90,width=0.5\textwidth]{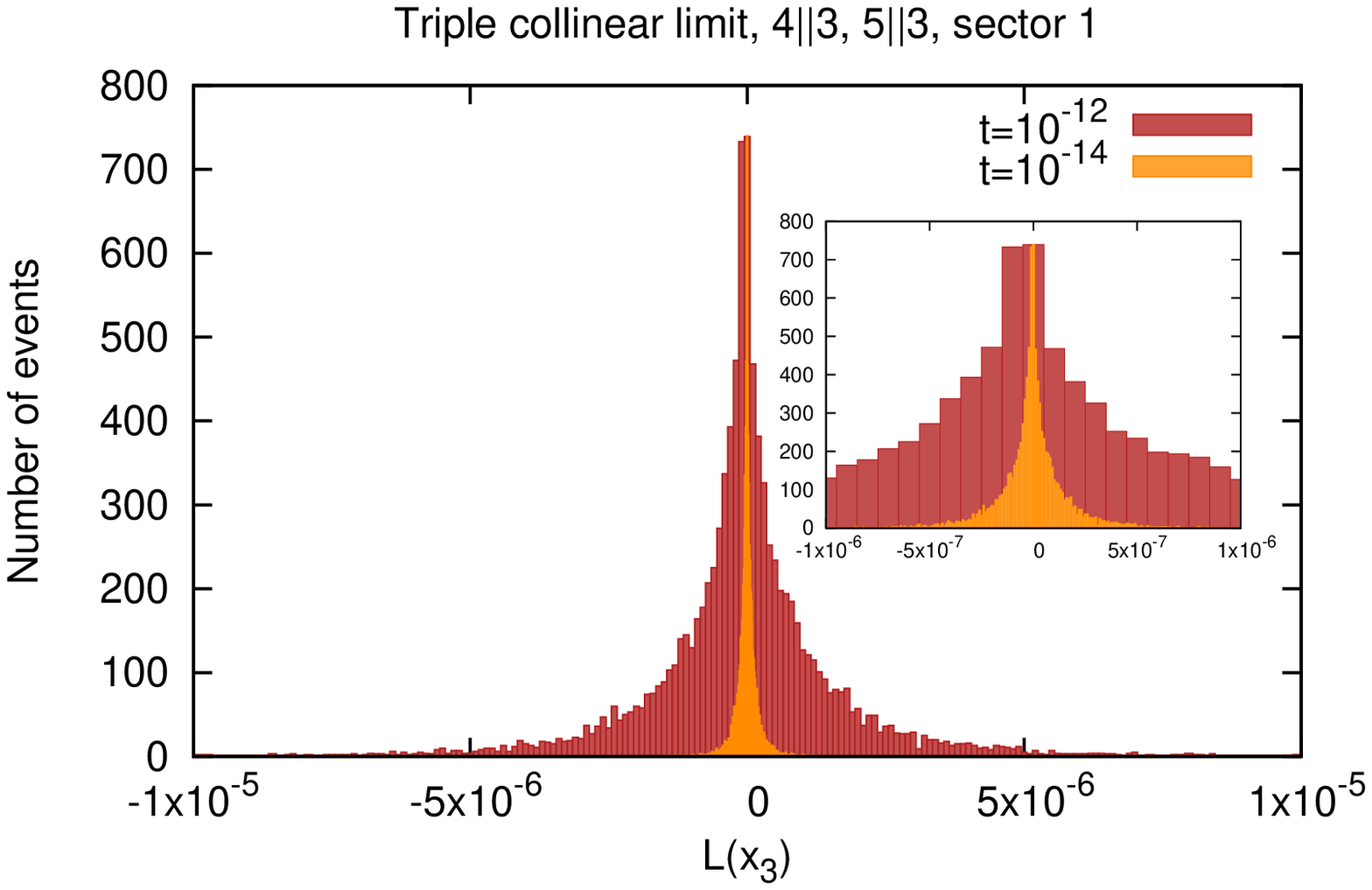}
\caption{Scaling behavior for soft (left) and collinear (right) 
double-real emission limits,
as obtained with our Fortran code in quadrupole precision.  
See the text for explanation. }\label{fig:rr_limit}
\end{figure}

For the double real  emission sectors, we consider 
fifteen  different limits, for example  
$L_{x_1}$, $L_{x_2}$, $L_{x_3}..$, $L_{x_1,x_2}...$, 
$L_{x_1,x_3,x_4},$ and check numerically how these functions 
approach zero. In particular, we know that 
all the soft limits should scale as  $t$, while  collinear limits 
should scale as $\sqrt{t}$. 
To illustrate this point   we plot distributions for the functions 
$L_{x_1}(t)$ and $L_{x_3}(t)$ in Fig.~\ref{fig:rr_limit}, 
for two sample sectors.  The function $L_{x_1}(t)$ describes 
the soft limit and the function $L_{x_3}$ describes the collinear limit. 
To obtain these plots, ten thousand $\vec x$ points were randomly 
generated and the two functions $L_{x_1}(t)$ and $L_{x_3}(t)$ were computed 
for two values of $t$ that differ either by one (soft) or two (collinear) 
orders of magnitude.  It is evident from Fig.~\ref{fig:rr_limit} that 
the widths of the resulting distributions scales with the parameter $t$ 
as expected.  We also note that, in case of the collinear limit, 
the quality of the distribution is very sensitive to the correct 
implementation of spin correlations.  In fact, by removing the 
spin-correlation part from collinear  splitting functions, we find  
$L_{x_3}(t) \sim {\cal O}(10^{-4}) $  independent of $t$ for $t \lsim 10^{-8}$. 

We note that we cannot follow the same strategy to check the  ${\cal O}(\ep)$ terms for  
lower-multiplicity amplitudes, since 
we do not have a computation of the $0 \to Hggggg$ amplitude beyond 
${\cal O}(\ep^0)$.  We can nevertheless check the consistency of 
our calculation and implementation 
by comparing  different limits against each other. 
In total, we consider 60 different combinations for 
all double-real sectors and check  
that each of them behaves in a way that is similar to what 
is shown  in Fig.~\ref{fig:rr_limit}, for $\ep = 0,1,2$.

\begin{figure}[!h]
\includegraphics[angle=-90,width=0.5\textwidth]{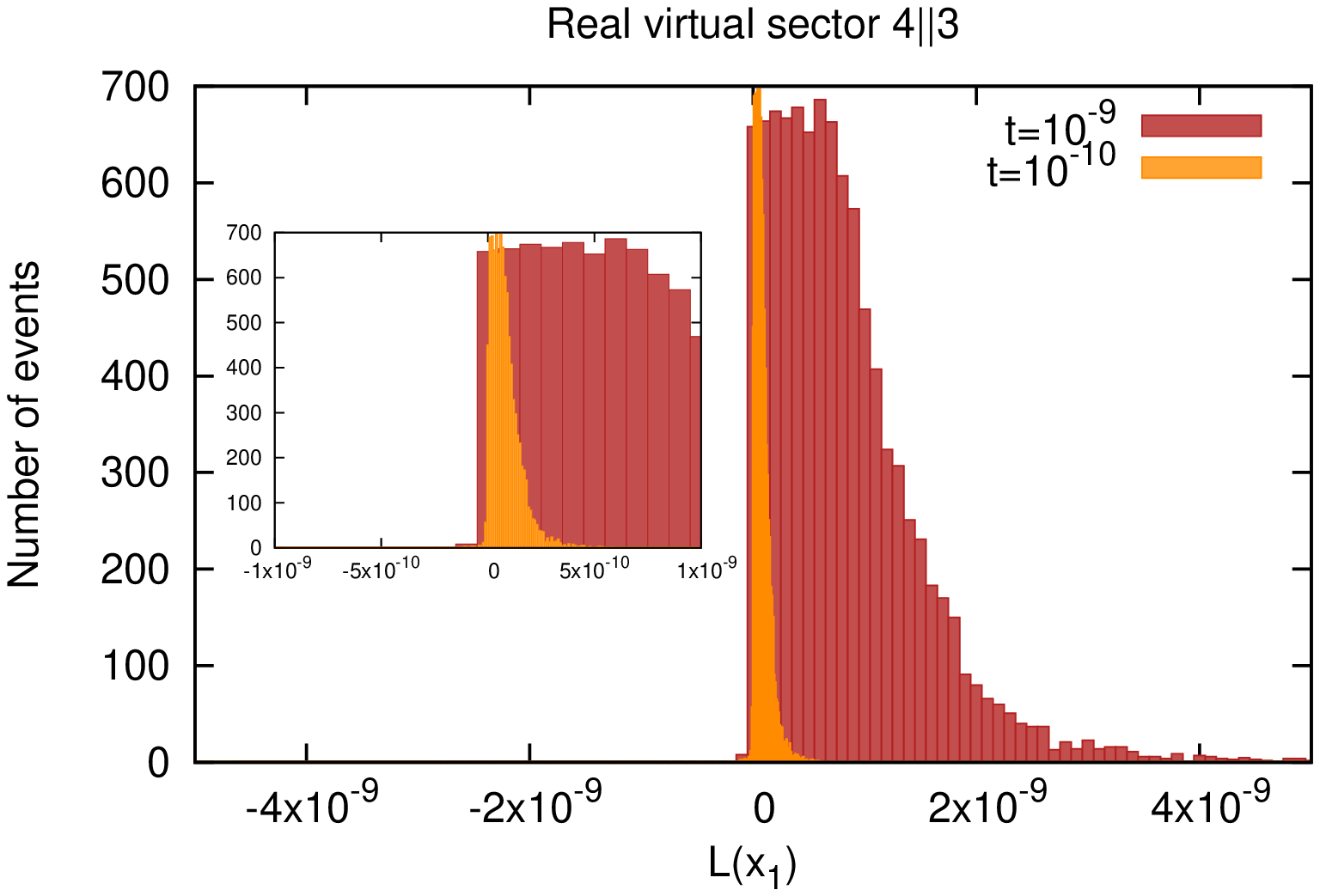}
\includegraphics[angle=-90,width=0.5\textwidth]{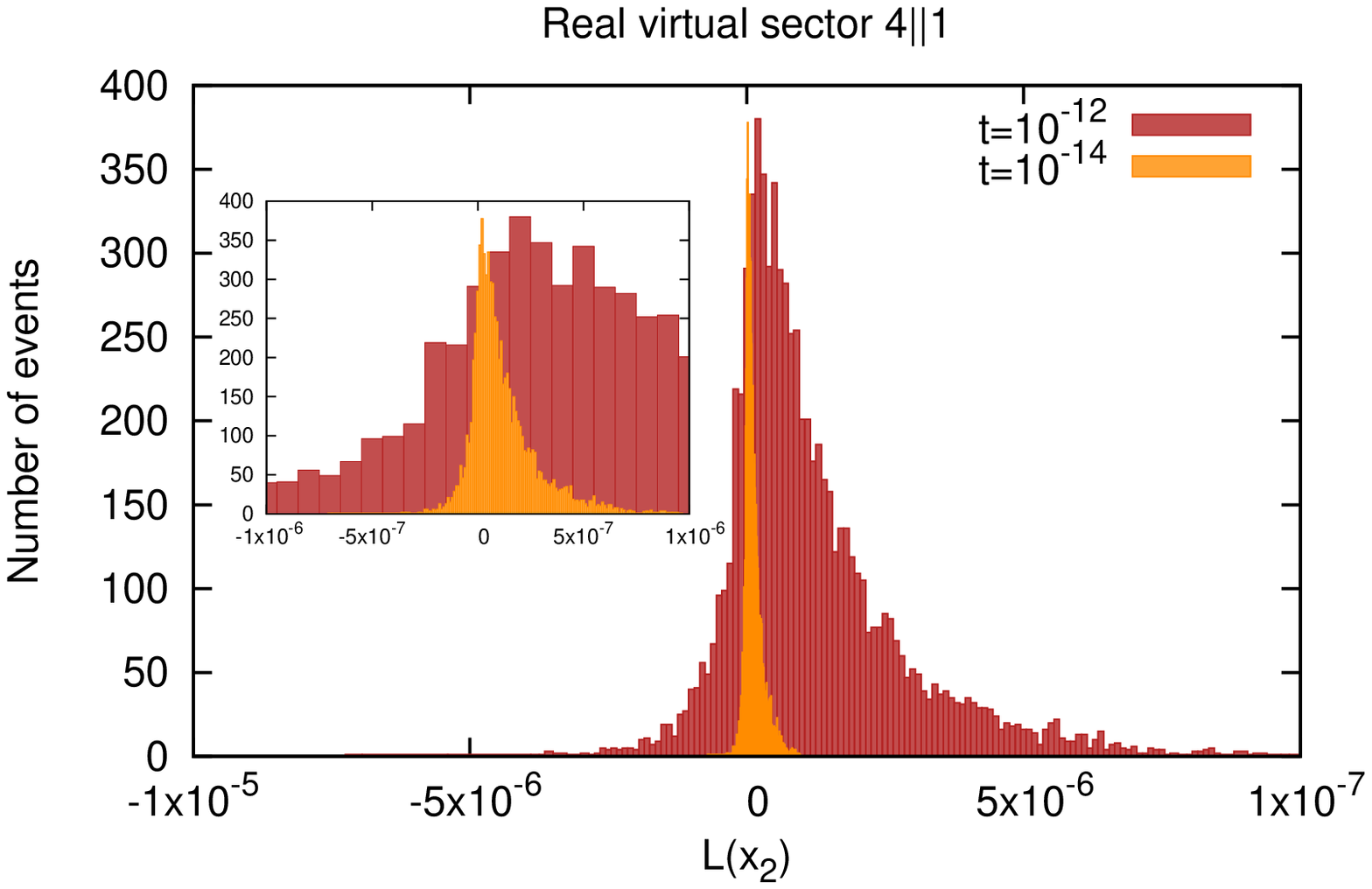}
\caption{
Scaling behavior for soft (left) and collinear (right) 
real-virtual  emission limits,
as obtained with our Fortran code in quadrupole precision.  
See the text for explanation.}\label{fig:rv_limit}
\end{figure}

To check the implementation of the real-virtual 
corrections, we need to modify the above strategy, since 
$F_{RV}(\vec x)$ is given by 
a linear combination of three functions with 
potentially logarithmically-singular coefficients, as shown in
Eq.~(\ref{eq_start0}). To probe soft and collinear limits 
in the real-virtual case, we define  two functions   
\be
 L_{1}(\ep,t) = 1 - 
\frac{ {\cal T}_\ep  \left [ F_{\rm RV}(t x_1,x_2,..
.) \right ]}{ {\cal T}_\ep  \left [ G_1(t,x_1,x_2,...) \right ] },
\;\;\;\;
L_2(\ep,t) = 1
- \frac{ {\cal T}_\ep \left [ F_{RV}(\ep,x_1,t x_2,...) \right ] }{
{\cal T}_\ep \left [ G_2(\ep,t,x_1,x_2,..) \right ] },
\label{eq_rv_lim}
\ee
where 
\be
\begin{split} 
& G_1(t,x_1,x_2,..) = F_1(0,x_2,...) 
+ F_2(0,x_2,...) \left[t^2 x_1^2 x_2\right]^{-\ep} 
+ F_3(0,x_2,...) \left[ t^2 x_1^2\right]^{-\ep},\\
& G_2(t,x_1,x_2,..) = 
F_1(x_1,0,...) + F_2(x_1,0,...) \left[t x_1^2  x_2 \right]^{-\ep}.
\end{split} 
\ee
The operator ${\cal T}_\ep$ in Eq.~(\ref{eq_rv_lim}) 
implies that the relevant term in the Laurent expansion in $\ep$ 
of the corresponding function should be taken.  For illustrative
purposes, we show distributions of $L_{x_1}(0,t)$ and $L_{x_2}(0,t)$ for one 
of the sectors in Fig.~\ref{fig:rv_limit}. Similar to the double real emission
case, we observe the ${\cal O}(t)$ scaling in  the soft limit and the
${\cal O}(\sqrt{t})$ scaling of the collinear limit. 

\begin{figure}[!h]
\begin{center}
\includegraphics[angle=-90,width=0.4\textwidth]{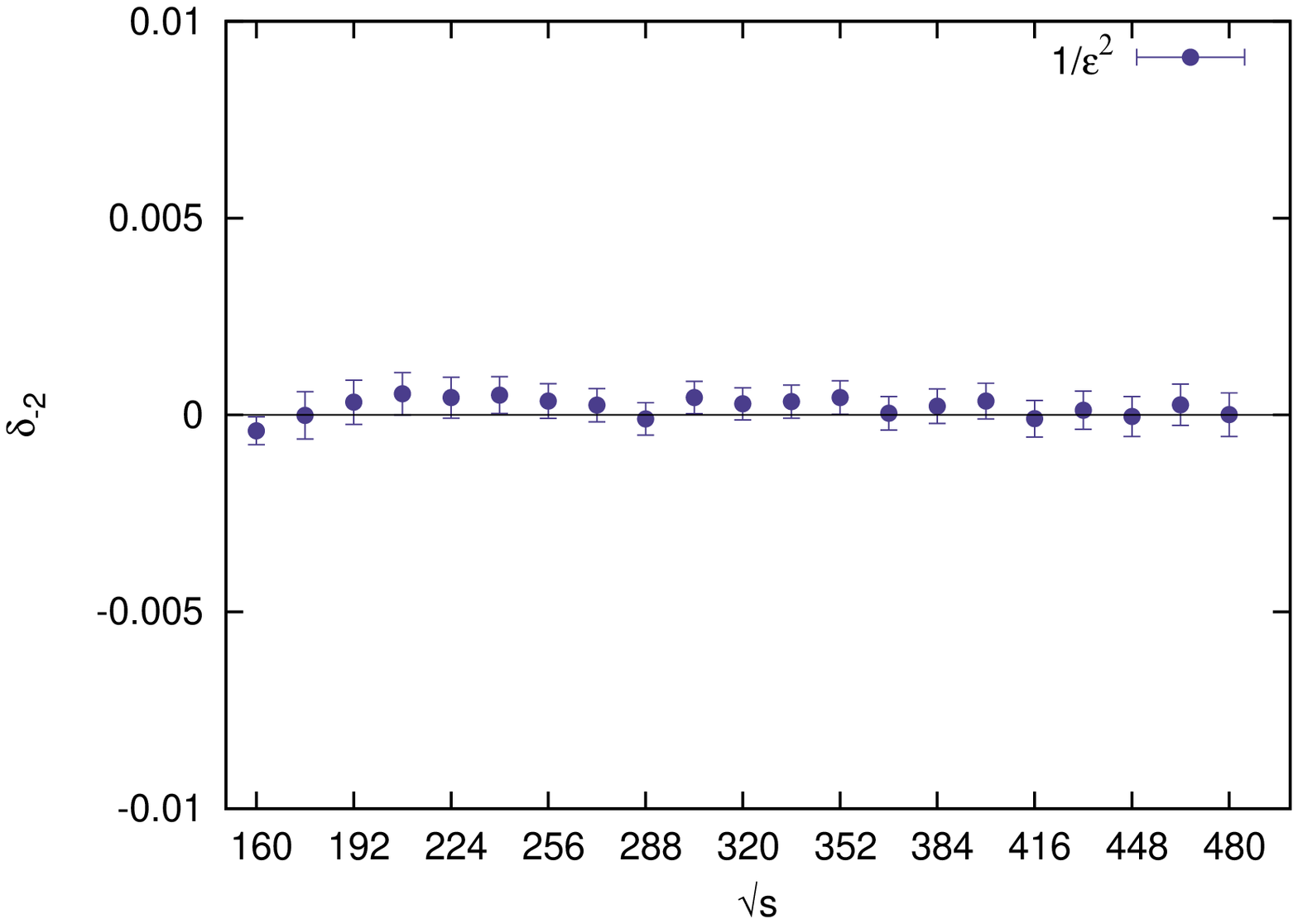}
\includegraphics[angle=-90,width=0.4\textwidth]{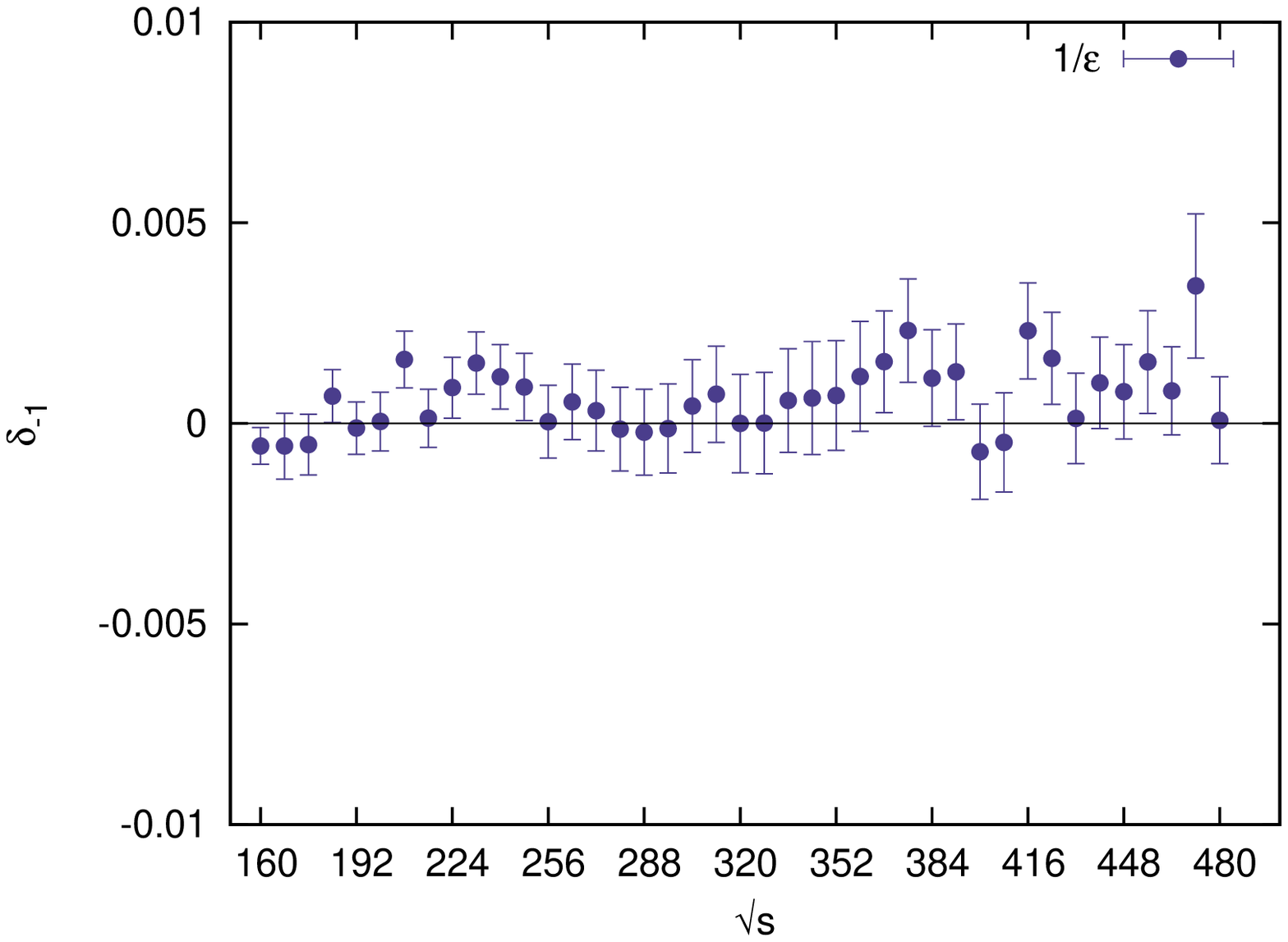}
\end{center}
\caption{Residuals of poles in $\ep$ for the total cross-section
as the function of partonic center-of-mass energy.  The left panel shows ${\cal O}(\ep^{-2})$, and the right panel
shows ${\cal O}(\ep^{-1})$.  
See the text for explanation. 
 }\label{fig:canc_poles}
\end{figure}

A further  check of the correctness of the calculation 
is provided by the  cancellation of poles. Singularities 
of double-real, real-virtual and double-virtual
 contributions start at ${\cal O}(\ep^{-4})$. Starting 
from order ${\cal O}(\ep^{-2})$,  collinear subtractions, renormalization 
and  contributions related 
to extra-dimensional components of the unresolved momenta
are required for the cancellation of poles. 
We note that  within our framework, we 
compute coefficients of the Laurent expansion in $\ep$ 
and check the cancellation of poles numerically.
To see how well this cancellation works, 
we compute the ratios  
\be
\delta_\ep = \frac{\sigma_{\rm RR}(\ep) + \sigma_{\rm RV}(\ep) 
+ \sigma_{VV}(\ep)
 + \sigma_{\rm conv}(\ep) + \sigma_{\rm renorm}(\ep) + \sigma_{d-4}(\ep)}
{ | \sigma_{\rm RR}(\ep)| + |\sigma_{\rm RV}(\ep)| + |\sigma_{VV}(\ep)|
 + |\sigma_{\rm conv}(\ep)| + |\sigma_{\rm renorm}(\ep)| + |\sigma_{d-4}(\ep)|}
\label{eq_ep_canc}
 \ee
at various orders in $\epsilon$.  In Eq.~(\ref{eq_ep_canc}), 
we account for  double-real, 
double-virtual, real-virtual contributions as well as 
 convolutions,  renormalization 
and the contribution due to extra-dimensional components of the 
unresolved gluon momenta.   We show $\delta(\ep)$ in 
Fig.~\ref{fig:canc_poles}  for $\ep = -2$ and $\ep = -1$. Interestingly, 
it appears  from Fig.~\ref{fig:canc_poles} 
that we loose  almost one order of magnitude in the quality 
of cancellation when we move from ${\cal O}(\ep^{-2})$ to 
${\cal O}(\ep^{-1})$. Nevertheless, at ${\cal O}(\ep^{-1})$ the cancellation 
is at the level of few per mille or better, which is acceptable.
Finally, we note that omission of extra-dimensional components in 
the momentum parametrization leads to residual 
non-cancellation of singularities at the level of $\delta_2
\sim 5 \times 10^{-3}$ 
and $\delta_1 \sim 2 \times 10^{-2}$, which is very large compared 
to values of $\delta$ that we observe in  Fig.~\ref{fig:canc_poles}.

As a final check of the calculation, we discuss the dependence of the result 
on  the 
renormalization and factorization scales. In this paper, we equate them and denote both by $\mu$.  We can compute  
the $\mu$-dependence of the cross-section either 
by introducing $\mu^{\ep}$ per coupling constant 
in the various elements of the  calculation in the standard way, 
or by solving the renormalization group equation  that 
follows from the fact that  convolution of the partonic cross-section 
with parton distribution functions is $\mu$-independent. The results 
of this computation can be found in 
Section~\ref{setup}. We have checked  that when the $\mu$-dependence 
is computed with our numerical code, the result 
agrees with the analytic computation based on renormalization group 
invariance.

\begin{figure}[!t]
\begin{center}
\includegraphics[width=0.65\textwidth]{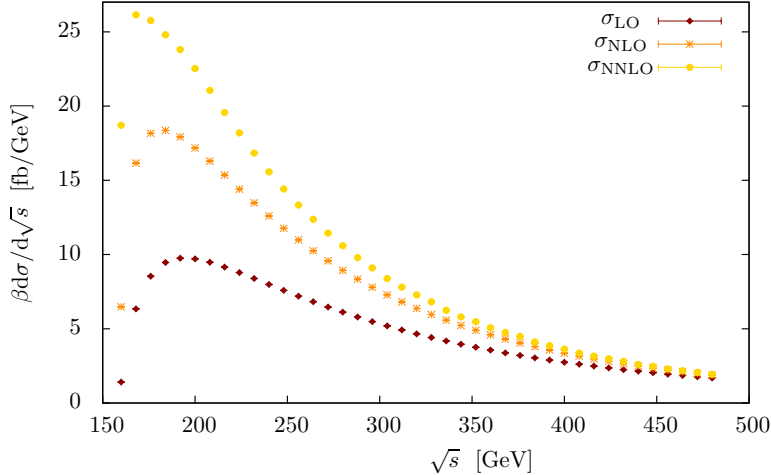}
\end{center}
\caption{Results for the product of partonic cross-sections 
$gg \to H+{\rm jet}$ and  parton luminosity in consecutive  orders in perturbative QCD
at $\mu_R = \mu_F = m_h = 125~{\rm GeV}$. 
See the text for 
explanation. 
 }\label{fig:xsect}
\end{figure}

We now present our results. We compute 
the hadronic cross-section for the production of the Higgs boson 
in association with a  jet  at the 8 TeV LHC 
through NNLO in perturbative QCD. 
We reconstruct jets using the $k_\perp$-algorithm
with $\Delta R = 0.5$ and $p_{\perp,j}=30~{\rm GeV}$. 
The Higgs mass is taken to be $m_H=125$~GeV
and the top-quark mass $m_t=172~$GeV. We use the 
latest NNPDF parton distributions~\cite{Ball:2012cx,Ball:2011uy} and 
numerical values of the strong coupling constant $\alpha_s$ 
at various orders in QCD perturbation theory as provided 
by the NNPDF fit. We note that in this case  $\as(m_Z)=[0.130,0.118,0.118]$ 
at leading,  next-to-leading  and next-to-next-to-leading 
order, respectively. 
We choose the central renormalization
and factorization scales to be $\mu_R=\mu_F=m_H$. In Fig.~\ref{fig:xsect} we show the partonic cross
section for $gg\to H+j$ multiplied by the gluon luminosity through NNLO in perturbative QCD
\be
\beta \frac{{\rm d}\sigma_{\rm had}}{{\rm d}\sqrt{s}} = \beta \frac{{\rm d}\sigma (s,\as,\mu_R,\mu_F)}{{\rm}d\sqrt{s}} \times
\mathcal L\lp\frac{s}{s_{\rm had}},\mu_F\rp,
\ee
where $\beta$ measures the distance from the partonic threshold,
\be
\beta = \sqrt{1-\frac{E_{th}^2}{s}},~~~~~~~ E_{th} = \sqrt{m_h^2 + p_{\perp,j}^2} + p_{\perp,j}\approx 158.55~{\rm GeV}.
\ee
The partonic luminosity $\mathcal L$ is given by the integral 
of the product of two gluon distribution functions 
\be
\mathcal L(z,\mu_F) = \int_z^1 \frac{{\rm d}x}{x} 
g(x,\mu_F) g\lp\frac{z}{x},\mu_F\rp.
\ee

It follows from Fig.~\ref{fig:xsect} that NNLO QCD corrections are significant in the 
region $\sqrt{s} < 500$~GeV. In particular, 
close to  partonic threshold $ \sqrt{s} \sim E_{th}$, radiative corrections are enhanced 
by threshold logarithms $\ln \beta$ that originate from the incomplete 
cancellation of virtual and real corrections.  There seems to be no significant enhancement 
of these corrections at higher  energies, where  the  NNLO QCD prediction for 
the partonic cross-section becomes almost indistinguishable from 
the NLO QCD one.  Note that we extend the  calculation of the NNLO partonic cross-section to 
$\sqrt{s} \sim 500~{\rm GeV}$ only.  From leading and next-to-leading 
order computations, we  know that by omitting the region
$\sqrt{s} > 500$~GeV,  we  underestimate the total cross-section by about 3\%.
To account for this in the NNLO hadronic  cross-section calculation, 
we  perform an extrapolation to higher energies constructed in such a way that when 
the same procedure  is applied to LO and NLO cross-sections, it gives results that agree well with the 
calculation without extrapolation.  The correction for the extrapolation  is included 
in the  NNLO QCD cross-sections results shown below. 

\begin{figure}[!t]
\begin{center}
\includegraphics[width=0.65\textwidth]{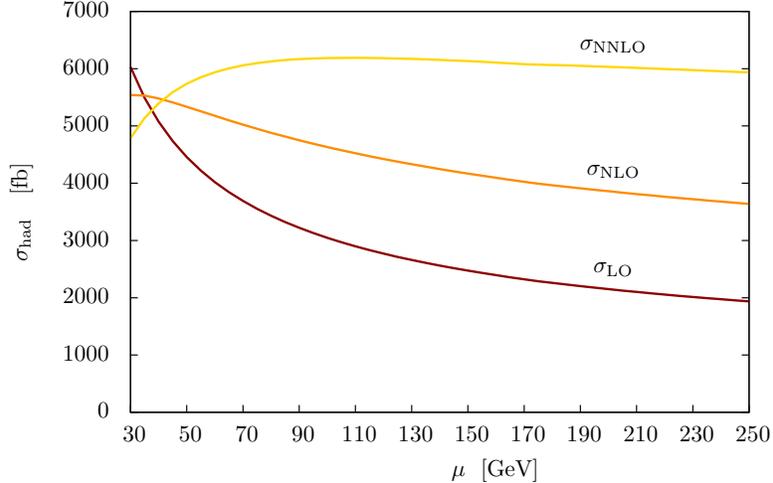}
\end{center}
\caption{Scale dependence of the hadronic cross section in consecutive  orders in perturbative QCD.
See the text for details. 
 }\label{fig:scale}
\end{figure}

We now show the integrated hadronic cross-sections for the 
 production of the Higgs boson in association 
with a jet at 8 TeV LHC in the all-gluon channel. 
We choose to vary the renormalization and factorization scale 
in the range $\mu_R=\mu_F = m_H/2,~m_H,~2m_H$. 
After convolution with the parton luminositites, we 
obtain\footnote{We checked our LO and NLO results  
against MCFM (gluons only), and
found agreement.}
\be
\begin{split} 
&\sigma_{\rm LO}(pp\to Hj) = 2713^{+1216}_{-776}~ {\rm fb}, \\
&\sigma_{\rm NLO}(pp\to Hj) = 4377^{+760}_{-738}~ {\rm fb}, \\
&\sigma_{\rm NNLO}(pp\to Hj) = 6177^{-204}_{+242}~ {\rm fb}.
\end{split} 
\ee
We note that NNLO corrections are sizable, as expected
from the large NLO $K-$factor, but the perturbative expansion shows  
marginal  convergence. We also evaluated PDFs error using the full set of NNPDF replicas,
and found it to be of order 5\% at LO, and of order 1-2\% at both NLO and NNLO, similarly
to the inclusive Higgs case~\cite{Ball:2012cx}. 
The cross-section increases by about sixty percent when we move from LO to NLO 
and by thirty percent when we move from NLO to NNLO.  It is also clear that 
by accounting for the NNLO QCD corrections we reduce the dependence on the renormalization 
and factorization scales in a significant way.  The scale variation of the 
result decreases from almost 50\% at LO, to 20\% at NLO, to less than 5\% at NNLO. 
We also note that a perturbatively-stable result is obtained for the scale 
choice $\mu\approx m_H/2$. In this case the 
ratio of the NNLO over the LO cross-section  is just 
$1.5$, to be compared 
with  $2.3$ for  $\mu=m_H$ and $3.06$ 
for $\mu=2m_H$, and the 
ratio of NNLO to  NLO is $1.2$.  It is interesting to point out that 
a similar  trend was observed 
in the calculation of higher-order QCD corrections 
to the Higgs boson production cross-section in gluon fusion. 
It has been pointed out that because of the rapid fall of the gluon PDFs, 
the production cross section is dominated by the 
threshold  region, thus making $\mu = m_H/2$ an excellent choice 
for the renormalization and factorization scales~\cite{Kramer:1996iq,kb}. 
The reduced scale dependence is also apparent from Fig.~\ref{fig:scale},
where we plot total cross-section as a function of the renormalization and
factorization scale $\mu$ in the region $p_{\perp,j}<\mu<2 m_h$.

Finally, we comment on the phenomenological 
relevance of the ``gluons-only'' results for cross-sections 
and $K$-factors that we reported in this paper. We note that 
at leading and next-to-leading order, quark-gluon 
collisions increase the $H+j$ production cross-section 
by about $30$ percent, for the input parameters that we use 
in this paper. At the same time, the NLO $K$-factors for the 
full $H+j$ cross-section are {\it smaller} by about $10-15$ percent 
than the ``gluons-only'' $K$-factors,  presumably because quark color charges 
are smaller than the gluon ones.  Therefore, 
we conclude that the gluon-only results 
can be used   for reliable phenomenological estimates 
of perturbative $K$-factors but adding quark channels will be essential 
for achieving precise results for the $H+j$ cross-section. We plan to 
return to this issue in the future.

\section{Conclusions}
\label{sec:conclusions}

In this paper we reported a calculation of the NNLO QCD corrections 
to the partonic process $gg \to H+{\rm jet}$.  This is one of the  first calculations
where NNLO QCD corrections are computed to a $2 \to 2$ process 
whose cross-section
depends on the 
implementation of the jet algorithm already  at leading order.  We believe that $gg \to Hg$ is 
a sufficiently typical process to expose 
all non-trivial features of a generic NNLO computation for 
a $2 \to 2$ process at a hadron collider. Indeed, we have used 
this process to show that the computational technique that 
we describe in this paper  can successfully deal with: 
\begin{itemize} 
\item a large number of contributing Feynman diagrams; 
\item colored particles in the initial and in the final state;
\item collinear subtractions and parton distribution functions;
\item all soft and collinear limits;
\item known helicity matrix elements; 
\item spin correlations;
\item a realistic jet algorithm.
\end{itemize}
The only ``non-generic'' feature  that we benefited from is a much 
simpler bookkeeping that is required for  
$gg \to Hg$  compared to the general case computation. 

We believe that the techniques  reported in this paper that  built 
upon earlier work described in Refs.\cite{Czakon:2010td,
Czakon:2011ve,Boughezal:2011jf},
allow computation of the NNLO QCD corrections to an arbitrary 
$2 \to 2$ process at hadron colliders provided that the corresponding 
two-loop matrix elements are available.  Since this is the case for 
most of the processes that are desirable to know at NNLO (c.f.  the ``NNLO wishlist'' in Ref.~\cite{huston}), 
our results open up a way to perform the required calculations.

On the other hand, it is not entirely clear to us how to extend 
the computational technology reported in this paper to 
make it  practically applicable to $2 \to n$, 
$n > 2$ processes. In this case, the problem is related to 
the ${\cal O}(\ep)$ parts of the amplitudes and the choice of 
extra-dimensional components to parametrize four-momenta of unresolved 
gluons. The point is that in the $2 \to 2$ process these details can 
still be dealt with by brute force, as we did in this paper, but 
for large $n$ this will be increasingly difficult to do.
Therefore, it is an interesting 
theoretical question to re-formulate this technique in such a way that
much of the irrelevant ${\cal O}(\ep)$ dependencies is  avoided. We 
hope to return to this point in the future. 

\bigskip
\noindent
{\bf {\Large Acknowledgments}} 
\medskip

\noindent
We thank T.~Gehrmann for clarifying to us some results in  Ref.~\cite{Gehrmann:2011aa}.
This research is partially supported by the US NSF under grants 
PHY-0855365  and PHY-1214000, by the U.S. Department of Energy, Division of High Energy 
Physics, under contract DE-AC02-06CH11357 and the grants DE-FG02-95ER40896 and DE-FG02-08ER4153, 
and by start-up funds provided by Johns Hopkins University. The research of K.M. is partially 
supported by Karlsruhe Institute of Technology
through a grant provided by its Distinguished Researcher Fellowship program.
Calculations reported in this paper were performed on the Homewood
High Performance Cluster of Johns Hopkins University or on computing resources provided by Argonne National Laboratory.

\appendix

\section{Appendix}
We report here the formulae
for the splitting functions and their convolution needed for the renormalization of parton distribution functions at NNLO, as described in Sec.~\ref{setup}.
\be
\begin{split} 
& P_{gg}^{(0)}(x) = 2 C_A \Bigg[  \frac{11}{12} \delta(1-x) + \left [ \frac{1}{1-x} \right ]_+ 
+ x(1-x) + \frac{1-x}{x} - 1 \Bigg ]
\\
& P_{gg}^{(0)} \otimes P_{gg}^{(0)} = C_A^2 \Bigg [  
\frac{22}{3} \left [ \frac{1}{1-x} \right ]_+
             + 8 \left [ \frac{\ln(1-x)}{1-x} \right ]_+
            + \left ( \frac{121}{36} - \frac{2}{3} \pi^2 \right ) \delta(1-x)
 \\
&             + \frac{4(x^4+3x^2+1-4x^3)}{x (x-1)}\ln(x)
           -\frac{8(2x-1+x^3-x^2)}{x}\ln(1-x)+\frac{2\left( 11 x^3-7 x^2-4 x-11 \right )}{3x}
\Bigg ]
\\
& P_{gg}^{(1)} =  C_A^2 \Bigg [ \lp\frac{8}{3} + 3\zeta_3\rp \delta(1-x) 
               + \lp\frac{67}{9}-2\zeta_2\rp \left [ \frac{1}{1-x} \right ]_+ 
+  \frac{4(x^2+x+1)^2}{x(1+x)} {\rm Li}_2(-x) 
\\
& + \frac{4 (3+2x^2+4x+2 x^3)}{2 (1+x)} \zeta_2
 + \frac{4 (x^2-x-1)^2}{2 (1-x^2)}\ln^2(x) - \frac{25}{18}- \frac{109}{18}x
\\
&  + \lp \frac{4(x^2+x+1)^2}{x (1+x)}\ln(1+x)
 - \frac{4 (x^2-x+1)^2}{x (1-x)}\ln(1-x)-\frac{75}{9}+\frac{33}{9}x-\frac{44 x^2}{3}\rp\ln(x) 
\Bigg ]
\end{split} 
\ee

\end{document}